\documentclass[%
 reprint,
 amsmath, amssymb,
 aps,
]
{revtex4-1}
\pdfoutput=1
\usepackage{graphics}
\usepackage{epsfig}

\usepackage{amssymb}
\usepackage{amsmath} 
\usepackage{color}

\usepackage{lineno}

\def \w {\omega}
\def \t {\theta}

\def \s {\sigma}

\def \< {\langle}
\def \> {\rangle}

\def \g {\gamma}

\newcommand{\rM}{\text}

\begin{document}
		
\title{A Phase Model Approach for Thermostatically Controlled Load Demand Response}

\author{Walter Bomela$^1$}
\email{wbomela@wustl.edu}

 \author{Anatoly Zlotnik$^2$}
\email{azlotnik@lanl.gov}

\author{Jr-Shin Li$^1$}%
 \email{jsli@wustl.edu}

\affiliation{$^1$Department of Electrical and Systems Engineering, Washington University, Saint Louis, MO 63130, USA\\
	$^2$Theoretical Division, Los Alamos National Laboratory, Los Alamos, NM 87545, USA\\}

		\begin{abstract}
			A significant portion of electricity consumed worldwide is used to power thermostatically controlled loads (TCLs) such as air conditioners, refrigerators, and water heaters.  Because the short-term timing of operation of such systems is inconsequential as long as their long-run average power consumption is maintained, they are increasingly used in demand response (DR) programs to balance supply and demand on the power grid.  Here, we present an \textit{ab initio} phase model for general TCLs, and use the concept to develop a continuous oscillator model of a TCL and compute its phase response to changes in temperature and applied power.  This yields a simple control system model that can be used to evaluate control policies for modulating the power consumption of aggregated loads with parameter heterogeneity and stochastic drift.  We demonstrate this concept by comparing simulations of ensembles of heterogeneous loads using the continuous state model and an established hybrid state model.  The developed phase model approach is a novel means of evaluating DR provision using TCLs, and is instrumental in estimating the capacity of ancillary services or DR on different time scales. 		We further propose a novel phase response based open-loop control policy that effectively modulates the aggregate power of a heterogeneous TCL population while maintaining load diversity and minimizing power overshoots.  This is demonstrated by low-error tracking of a regulation signal by filtering it into frequency bands and using TCL sub-ensembles with duty cycles in corresponding ranges.  Control policies that can maintain a uniform distribution of power consumption by aggregated heterogeneous loads will enable distribution system management (DSM) approaches that maintain stability as well as power quality, and further allow more integration of renewable energy sources.

		\end{abstract}
		\maketitle
		
		
	
	
	
	\section{INTRODUCTION}
		
	Significant efforts have been made in recent years to reduce the greenhouse gas emissions caused by fossil-fueled power plants throughout the world.  In the United States, many states have been adopting or increasing their renewable energy generation portfolio standards \cite{Kara14}.  The increasing penetration of such renewable energy sources (RESs) affects power quality on electric distribution systems and complicates load balancing in power systems \cite{YAN2017241}.  These issues compel the development of new approaches to regulate the inherently fluctuating and uncontrollable power outputs of RESs \cite{Kara15, Meyn15}.
	
Demand response (DR) programs enable electricity users to adjust their consumption in response to energy prices or incentive payments \cite{BEHBOODI2017, CHEN2017}, and thus provide significant capability to balance supply and demand on the power grid. Indeed, demand-side management technology has been under development since the late 1970's and early 1980's \cite{Bischke85}, with consideration of diverse objectives that include peak shaving, valley filling, and strategic energy conservation \cite{Delgado85}.  With the advent of smart grid technologies and the integration of RESs in distribution systems around the world, system operators will now have to rely on DR and ancillary services (AS) to balance supply and demand of energy more than ever before.  It is therefore imperative to develop technologies that can exploit the resources available for DR to the fullest extent.
	
A comprehensive review  on the current challenges and barriers to a full   deployment of demand response programs is given in \cite{NOLAN20151}. One important observation made in the review is that, despite receiving information about their energy consumptions, the majority of the participants in a DR study continued with their everyday routines and habits. This kind of consumer behavior will diminish the potential of DR.  In order to maximize DR capabilities, it may be better to remove human decisions from the loop.  This could be achieved by using a transactive control paradigm, which is a popular direct load control approach in which customers have agreed to allow the power company to control some of their flexible loads \cite{CHEN2017}. This control approach reacts faster to market price fluctuations and enables TCLs to participate in the real-time retail electricity market \cite{BEHBOODI2017}.

	The integration of RESs into the power grid reduces the inertia of the system, which poses a challenge for automatic generation control (AGC), which is used to control frequency by balancing supply and demand. Several studies have examined ways to utilize electric vehicles (EVs) and battery energy storage systems (BESSs) to provide frequency regulation reserves \cite{SIANO2014461, pavic2016low}. For instance, \cite{TENG2017353} developed a techno-economic evaluation framework to quantify the challenges and assess benefits to primary frequency control by using EV batteries. Although BESSs can provide primary frequency control, the cost of deployment in large capacities remains prohibitively high \cite{MALIK2017}. Hence, the focus has been turned to TCL loads  such as refrigerators \cite{LAKSHMANAN2016470, LAKSHMANAN2016705},  heating ventilation and cooling (HVAC), and electric water heaters (EWH) that have been shown to be suitable for providing ancillary services to the grid \cite{ ZHOU2017456, Ghaffari15}.

	In this paper, we consider the direct control of TCLs through thermostat set-point changes for DR provision. Although we consider AC systems, the modeling and control approach presented herein is applicable to any general TCL.  In the recent past, various modeling and control methods for TCLs were developed, and experiments were conducted to demonstrate the ability of TCLs to provide ancillary services.  Field experiments conducted using domestic refrigerators in \cite{LAKSHMANAN2016705} and \cite{LAKSHMANAN2016470} aimed at quantifying the flexibility of household TCLs, and the computational resource constraints for the control of large TCL populations.  The frequency controller in \cite{LAKSHMANAN2016470} switches the TCLs OFF one by one based on the ability of each refrigerator to stay OFF longer. In \cite{LAKSHMANAN2016705}, the same authors introduced a delay in their control scheme to improve the performance of the controller, which reduced the power overshoot. They noted that withdrawing a large number of loads abruptly produced instability and caused the loads to synchronize their duty cycles.
	In \cite{MALIK2017}, such synchronization of TLCs is avoided by randomizing the parameters in the control scheme.  This  approach consisted of dividing the compressor cycle into four different states based on the compartment temperature.
	Their control architecture was based on the utility sending a regulation signal to Cooperative Home Energy Management (CoHEM) systems located  at distribution transformers, with each CoHEM then sending different control signals to home energy management systems that control the refrigerators.
	The investigation in \cite{ma2017switched} combined three different control protocols to improve the accuracy in tracking an AGC signal. The controller switched between a temperature priority based control, a sliding mode control, and a two-stage regulation control.
	A two-level scheduling method intended to facilitate the scheduling of flexible TCLs in the intra-day electricity market was proposed in \cite{ZHOU2017456}. However, the performance of this method deteriorates with parameter heterogeneity, and worsens as forecast uncertainty increases.
	Control approaches based on model predictive control techniques have also been considered. In \cite{BAETEN2017184}, a multi-objective model predictive control strategy for residential heating with heat pumps was proposed. This approach takes into account the users' energy cost, the environmental impact, and expansion of electricity generation capacity. In addition, it considers detailed models for heat pump and thermal energy storage, and accounts for the feedback effect of individual controllers on electricity generation.  This level of detail will make scaling the approach to a large TCL population problematic.

One very important question that was posed in \cite{ARYANDOUST2017749}, and that still needs to be accurately answered, is \emph{how large is the potential of DR and on which time scale can DR be the most effective?} This is not a particularly easy question to answer.
	In this paper we propose a novel procedure for evaluating the maximum capacity and the appropriate time scale for DR provision by a collection of TCLs. Based on phase reduction theory \cite{izhikevich2007dynamical}, this approach can evaluate the capacity of a TCL population in different bandwidths.
	In \cite{barooah2015spectral}, a theoretical upper bound on the capacity of TCLs to provide ancillary services as a function of frequency was proposed. The capacity-bandwidth constraints was derived based on standard linear dynamical system models. The proposed bound indicates that for a given TCL population with a fixed time constant, the capacity for AS provision is inversely proportional to the frequency of the regulation signal to track.
	Our method supplements this observation by further showing that the capacity does not monotonically decreases with frequency. 	
	Indeed, while the capacity monotonically decreases for a regulation signal with frequency below the mean natural frequency of a given TCL population, it then increases and decreases again in some frequency bandwidths higher than the TCLs' mean frequency.
	This is particularly significant because this theory can be used to analyze different TCL populations and classify them based on their capacities to provide AS in different bandwidths (time scales).

	Unwanted synchronization of TCL duty cycles by the control policy is most likely the main factor that limits the time scale and capacity of ancillary services that a given population can provide.
	This phenomenon has motivated the research and development of control policies  such as the safe control protocols \cite{sinitsyn2013safe,mehta2014safe}, which aim to minimize unwanted power oscillations in response to pulse-like changes of the set-point temperature. One protocol allows the TCL to stay in its current state until the temperature hits one of the transition points, then starts following a new pair of deadband limits \cite{sinitsyn2013safe}, and the other adds a delay of M-minutes before changing the status of the TCL  \cite{mehta2014safe}.
	Other control strategies with similar goals consist of turning ON/OFF some TCLs in the OFF/ON stack based on their priority measure \cite{Hao15}, \cite{Vrettos2012load}.  The priority measure is defined as the distance from the switching boundary and hence, if the TCLs' aggregate power needs to be reduced, the TCLs in the ON stack closer to the switching boundary will be turned OFF first. Conversely, if the aggregate power needs to be increased to follow an increase in the generated power, the TCLs in the OFF stack closer to the switching boundary will be turned ON first.

	In an effort to further the understanding of the oscillatory behavior of the aggregate power consumption, the damping of oscillations in aggregate power was characterized as a function of parameter heterogeneity by exploiting the similarities that exist between a population of mass-springs systems and an ensemble of  TCLs \cite{docimo2017demand}.
	Concurrently, the dependence of the mixing rate of the population on the model parameters was characterized in \cite{chertkov2017ensemble}.

	 To summarize, the work in this paper provides an novel approach for evaluating DR capacity and time scale using a novel phase model representation of TCLs. The evaluation is conducted by computing the entrainment regions also known as Arnold tongues \cite{izhikevich2007dynamical}.
	To obtain the phase model,  we started by first extending the common hybrid-state model of TCL dynamics to a neuroscience-inspired representation in the form of a continuous two-dimensional system, then applying a widely-used phase reduction computation \cite{Nakao15}.
	The phase model is a one-dimensional system with scalar state, and its simplicity has made it one of the most popular models for studying oscillatory systems, including power grids \cite{Dorfler11} and neural oscillators \cite{Ermentrout96}, with particular advantages for control design in the presence of parameter uncertainty \cite{Zlotnik12}. The second novel contribution of this work is the proposed phase response based open-loop control policy, which exploits the natural diversity that exists amongst the phase response functions of TCLs. 	
	This approach is promising for developing hierarchical grid control architectures that maintain power quality on distribution systems with large penetration of RESs while providing load balancing services to regional transmission operators (RTOs).  To the best of our knowledge, this paper is the first to consider the modeling and control of TCLs using the phase model reduction approach.

	The rest of the paper is structured as follows.  An \emph{ab initio} scalar phase model of a basic hybrid system representation of a single TCL is derived in Section \ref{sec:Basic_D_mdl}. Then, a continuous two-state TCL model and its phase-reduced representation are derived in Sections \ref{sec:2D_mdl} and \ref{sec:TCL_Phase_mdl}.  Monte Carlo simulations of an example TCL described using the two models are then compared in Section \ref{sec:Sim_mdl}.
	After this model validation, in Section \ref{sec:ControlPolicy} we analyze the synchronization properties of TCL ensembles in order to enable the evaluation of DR capacity and time scales, and explore a phase response based open-loop control policy whose  efficacy for controlling the aggregate power of TCL ensembles is demonstrated by tracking a regulation signal from the Balancing Authority (BA) decomposed into power spectrum components.
	A discussion of applications and promising future research directions is given in Section \ref{sec:Conclusion}.
	
	\section{Thermostatically Controlled Load Models}
	\label{sec:TCL_mdl}
	
	\subsection{One-Dimensional Hybrid Model}
	\label{sec:1D_mdl}
	
	The dynamics of the internal temperature $\theta(t)$ of a house equipped with an air-conditioning system is often described using a simple hybrid-state model \cite{Ihara81}.  The model describes how an AC unit regulates the average temperature by means of a thermostat and a relay with state $s(t) \in \{0,1\}$ \cite{Perfumo12}.  The hybrid state model describing the evolution of the internal temperature $\theta(t)$ is given by
	\begin{equation}\label{eq:Temperature}
	\dot{\t}(t) = -\frac{1}{RC} [ \t(t) - \t_a + s(t) PR ],
	\end{equation}
	\begin{equation}\label{eq:switch}
	s(t) = \left\{ \begin{array}{lll}
	0 & \mbox{if $\t(t) < \t_{\min}$}\\
	1 & \mbox{if $\t(t) > \t_{\max}$}\\
	s(t) & \mbox{otherwise}, \end{array} \right.
	\end{equation}
	where $\t_a$ represents the ambient temperature, $P$ is the average energy transfer rate of the TCL in the ON state, $C$ and $R$ are the thermal capacitance and resistance of the house, respectively. The minimum and maximum temperature of the TCL are $\t_{\min} = \t_s-\delta/2$ and $\t_{\max} = \t_s+\delta/2$, respectively, with $\t_s$ the thermostat  temperature setpoint and  $\delta$ the deadband.
	Consider a population of $N$ TCLs with temperature states $\t_i(t)$ for $i\in \{1,\cdots, N \}$ that evolve according to \eqref{eq:Temperature}.  Then the aggregate power $P_{agg}(t)$ drawn by all $N$ TCLs is \cite{Bashash11}
	\begin{equation}\label{eq:P_aggr}
	P_{agg}(t) = \sum_{i=1}^N \frac{1}{\eta_i} s_i(t) P_i,
	\end{equation}
	where $\eta_i$ is the coefficient of performance of each TCL unit.
	
	\subsection{Ab Initio Deterministic TCL Phase Model}
	\label{sec:Basic_D_mdl}
	
	We formulate a basic model of a TCL as a deterministic switched oscillating system, whose state is given by the temperature $\theta\in[\theta_{\min},\theta_{\max}]$ and the switching function $s\in\{\rM{OFF, ON} \}$.  Suppose without loss of generality that the system is an air conditioner that cycles through a duty cycle to maintain a temperature $\theta_{s}=\frac{1}{2}(\theta_{\max}-\theta_{\min})$ while remaining within the deadband $\delta=\theta_{\max}-\theta_{\min}$.  We denote the beginning of the cycle as the point $(\theta,s)=(\theta_{\max},\rM{ON})$ just after the unit has turned on.  The state of the unit changes according to $\dot{\theta}=r_-<0$ when the unit is at $s=\rM{ON}$, then it switches to $s=\rM{OFF}$ when the state reaches $\theta=\theta_{\min}$.  Then, the state of the unit changes according to $\dot{\theta}=r_+>0$ until it reaches $\theta=\theta_{\max}$, and the unit turns on again.
	Let us call the length of time when the unit is ON $T_{\rM{ON}}$, and the length of time when the unit is OFF $T_{\rM{OFF}}$, so that the period of oscillation is $T=T_{\rM{ON}}+T_{\rM{OFF}}$.
	
	\begin{figure}[!h]
		\centering
		\includegraphics[width=.9\linewidth]{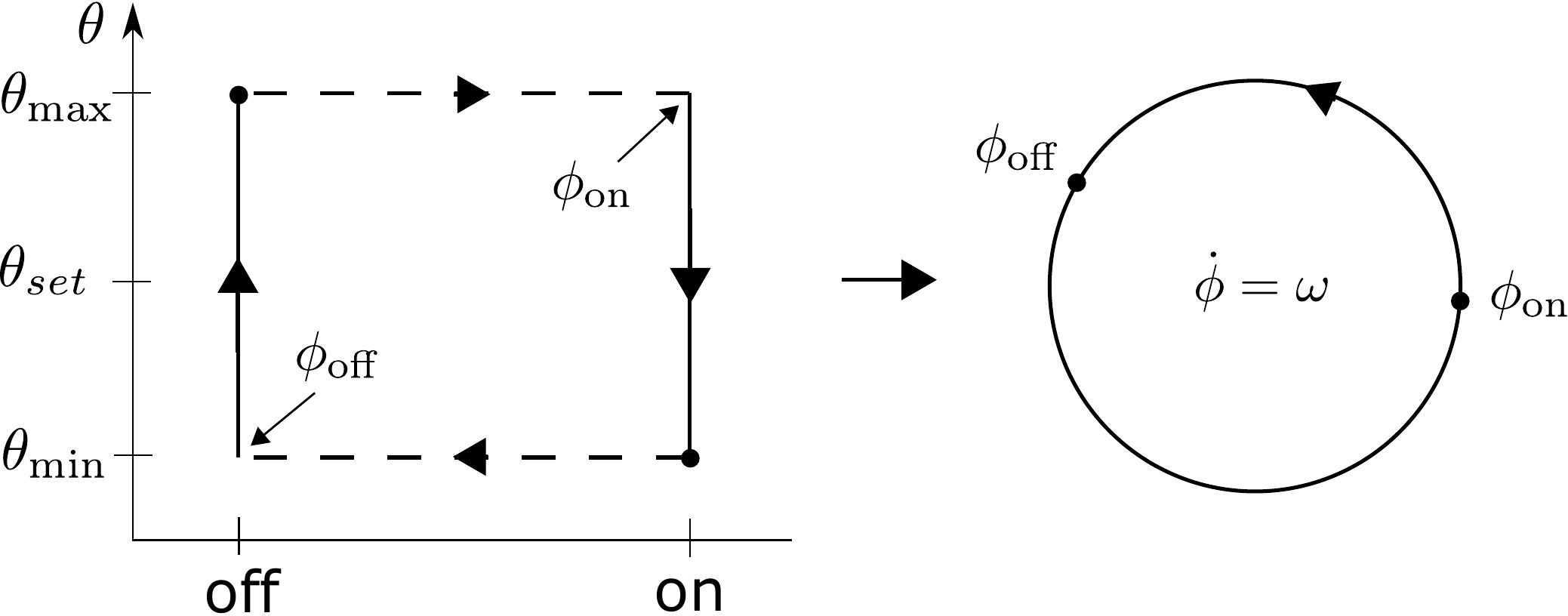}
		\caption{Phase model of a switched oscillating system.} \label{fig:duty0}
		\vspace{-1ex}
	\end{figure}
	
	Our goal is to map this behavior to a phase model, as illustrated in Fig. \ref{fig:duty0}.  Such models are desirable because the homogeneous dynamics are linear and scalar.  The state is represented by a scalar phase $\phi$, which advances linearly with time according to a frequency $\omega=2\pi/T$.  For an unforced system, this yields a simple solution $\phi=\omega t$ $(\rM{mod} 2\pi)$.  Thus, we map the state point $(\theta_{\max},\rM{ON})$ where the unit turns ON to the phase point $\phi=\phi_{\rM{ON}}\equiv 0$, and the state point $(\theta_{\min},\rM{OFF})$ where the unit turns OFF to the phase point $\phi=\phi_{\rM{OFF}}$.  This yields a continuous representation of the switched system.
	
	To complete the picture, we first determine the phase $\phi_{\rM{OFF}}\in[0,2\pi)$ when the unit switches OFF.  If $r_+=r_-$, then it is straightforward to show that $\phi_{\rM{OFF}}=\pi$.  However, most units will have different values of $r_+$ and $r_-$, which will also depend on other factors such as the ambient temperature.  Suppose then that $r_+\neq r_-$, and let us denote the rate ratio, which is equivalent to the duty ratio, by
	\begin{align} \label{eq:rateratio0}
		\gamma &= -\frac{r_-}{r_+} = \frac{T_{\rM{OFF}}}{T_{\rM{ON}}}.
	\end{align}
	It is straightforward to show that these ratios are equivalent.  Integrating the $s=\rM{ON}$ dynamics $\dot{\theta}=r_-$ from $t=0$ to $t=T_{\rM{ON}}$ yields
	\begin{align} \label{eq:rateratio1a}
		r_-T_{\rM{ON}} &=\theta_{\min}-\theta_{\max}=-\delta,
	\end{align}
	and integrating the $s=\rM{OFF}$ dynamics $\dot{\theta}=r_+$  from  $t=T_{\rM{ON}}$ to $t=T=T_{\rM{ON}}+T_{\rM{OFF}}$ yields
	\begin{align} \label{eq:rateratio1b}
		r_+T_{\rM{OFF}} &=\theta_{\max}-\theta_{\min}=\delta.
	\end{align}
	Then, equation \eqref{eq:rateratio0} follows directly from \eqref{eq:rateratio1a} and \eqref{eq:rateratio1b}, and we then have $\gamma>0$, and $T_{\rM{OFF}}=\gamma T_{\rM{ON}}$.  We can then compute $T_{\rM{ON}}$ and $T_{\rM{OFF}}$ according to
	\begin{equation*} \label{eq:dutytime}
		\begin{split}
			T_{\rM{ON}} &= T - T_{\rM{OFF}} = T - \gamma T_{\rM{ON}}
			= \frac{1}{1+\gamma} T = \frac{r_+}{r_+-r_-} T, \\
			T_{\rM{OFF}} &= T - T_{\rM{ON}} = T - \frac{1}{1+\gamma} T
			= \frac{\gamma}{1+\gamma} T = \frac{-r_-}{r_+-r_-} T.
		\end{split}
	\end{equation*}
	It follows that the switch-OFF  phase is given by
	\begin{equation*} \label{eq:phaseoff}
		\phi_{\rM{OFF}}  = \frac{2\pi}{T} T_{\rM{ON}} = 2\pi\frac{r_+}{r_+-r_-} = 2\pi\frac{T_{\rM{ON}}}{T_{\rM{ON}}+T_{\rM{OFF}}}.
	\end{equation*}
	
	It is straightforward to show that the parameters in the hybrid-state model given in equations \eqref{eq:Temperature}-\eqref{eq:switch} are related to the phase model parameters by
	$$ T_{\rM{OFF}} = RC \ln \left( \frac{ \t_{\rM{a}} - \t_{\rM{min}}  }{ \t_{\rM{a}} - \t_{\rM{max}} }\right) \text{,  }	$$	
	$$T_{\rM{ON}} = RC \ln \left( \frac{ \t_{\rM{max}} - \t_{\rM{a}} + PR }{ \t_{\rM{min}} - \t_{\rM{a}} + PR } \right),$$
	with the period given by
	\begin{align} \label{eq:T}
		T &= RC \ln \left( \frac{ (\t_{\rM{a}} - \t_{\rM{min}}) (\t_{\rM{max}} - \t_{\rM{a}} + PR)  }{ (\t_{\rM{a}} - \t_{\rM{max}}) (\t_{\rM{min}} - \t_{\rM{a}} + PR )  } \right).	
	\end{align}
	Hence, we can express the switch-off phase by
	\begin{align*}
		\phi_{\rm{OFF}} &= 2\pi \frac{\ln \left( \frac{ \t_{\rM{max}} - \t_{\rM{a}} + PR }{ \t_{\rM{min}} - \t_{\rM{a}} + PR } \right)}{\ln \left( \frac{ (\t_{\rM{a}} - \t_{\rM{min}}) (\t_{\rM{max}} - \t_{\rM{a}} + PR)  }{ (\t_{\rM{a}} - \t_{\rM{max}}) (\t_{\rM{min}} - \t_{\rM{a}} + PR )  } \right)}.
	\end{align*}
	Note that to control the average power of a TCL, one needs to modulate the duty cycle of the power utilization. This is equivalent to controlling the phase $\phi_{\rM{OFF}}$ of the phase model illustrated in Fig. \ref{fig:duty0} assuming for instance that $\phi_{\rM{ON}}=0$  is the reference phase.
	
	\subsection{Continuous Representation of Switching Dynamics}
	\label{sec:2D_mdl}
	
	We now present a two-dimensional continuous state-space model that approximates the dynamics of the internal temperature of a TCL unit and the thermostat switching.    This is an important step that precedes the derivation of the complete phase model, which will require the computation of the phase response curve (PRC) of the continuous oscillator model.    Subsequently, we will examine the phase response of the temperature dynamics to control action applied to the TCL.
	
\begin{figure}[!h]
		\centering
		\includegraphics[width=1\linewidth]{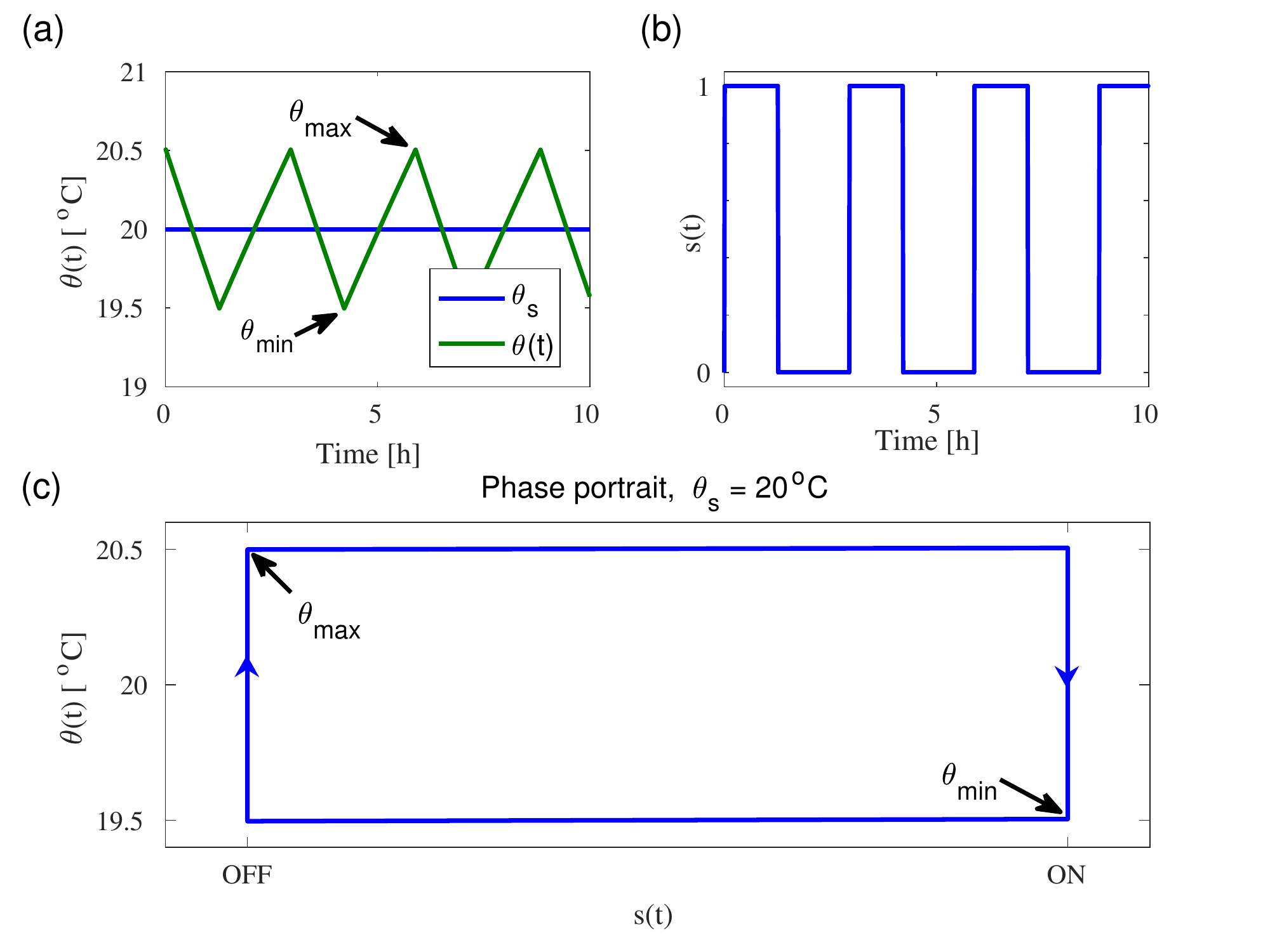}
		\caption{Simulation of the hybrid model \eqref{eq:Temperature}-\eqref{eq:switch} with the parameters in Table 1 and the deadband $\delta=1.0^{\circ}$C. (a) Time evolution of the temperature $\t(t)$ around the setpoint $\t_s = 20^{\circ}$C. (b) Thermostat switching function $s(t)$. (c) Phase portrait. }
		\label{fig:Fig1_Temp_switching}
		\vspace{-1ex}
	\end{figure}

	Consider the evolution of the temperature $\t(t)$ (see Fig. \ref{fig:Fig1_Temp_switching}a) described by  \eqref{eq:Temperature}, and the corresponding phase portrait in Fig. \ref{fig:Fig1_Temp_switching}c,  simulated with the parameters provided in Table 1.  Observe that the unperturbed behavior of a TCL  is similar to that of an oscillator with a stable limit cycle.
	
	\begin{table}[!h]
		\renewcommand{\arraystretch}{1.3}
		\caption{Nominal TCLs parameter values \cite{Callaway09}}
		\label{tab:example}
		\centering
		\begin{tabular}{c|c|c}
			\hline
			Parameter  &  Meaning	&	Value\\
			\hline
			\hline		
			$\t_s$   &   temperature setpoint	&	$20^{\circ}$C\\
			$\t_a$   &   ambient temperature 	&	$32^{\circ}$C\\
			$\delta$   &   thermostat deadband 	&	$0.5^{\circ}$C\\
			$R$   &   thermal resistance 	&	$2^{\circ}$C/kW\\
			$C$   &   thermal capacitance 	&	$10$ kWh/$^{\circ}$C\\
			$P$   &   energy transfer rate 	&	$14$ kW\\
			$\eta$   &   coefficient of performance 	&	$2.5$\\
			\hline
		\end{tabular}
	\end{table}
	
	The hybrid-state nature of the system described by \eqref{eq:Temperature} and \eqref{eq:switch} is due to the thermostat switching function $s(t)$ that transitions between $1$ and $0$ states.
	Therefore, modeling the behavior of a TCL using continuous states requires a continuous approximation of the switching function $s(t)$ and its dynamics.  Ideally, the evolution of the temperature and the continuous switching function could be represented using a system of two coupled differential equations that has a stable limit cycle similar to the one shown in Fig. \ref{fig:Fig1_Temp_switching}c.
	Our motivation for the proposed model is the Van der Pol oscillator, which is a simple model of the limit cycle observed in circuits with vacuum tubes. More so, a similar phase portrait is observed in the FitzHugh-Nagumo model, which is a simple mathematical description of the firing dynamics of a neuron \cite{Nakao15}.
	Inspired by these examples, we propose the continuous-state TCL model given by
	\begin{equation}\label{eq:Continuous_mdl_v1}
	\begin{split}
	\dot{x}(t) &= \mu \left( (\frac{\delta}{2}+\s) x - \frac{x^3}{3} +\t - \t_s \right),\\
	\dot{\t}(t) &= -\frac{1}{RC} \left( \t - \t_a +\bar{s}(t) PR \right),
	\end{split}	
	\end{equation}
	where $x(t)$ is the state variable of the switching function, $\t(t)$ the internal temperature, and $\bar{s}(t)$ is an approximation of the ideal switching function $s(t)$. The parameter $\s$ was introduced to compensate for the reduction of the effective deadband in \eqref{eq:Continuous_mdl_v1}. The constant $\mu$ is a damping parameter that controls the oscillation frequency for a fixed time constant $\tau=RC$
	as well as the shape of the phase portrait. Just like for the Van der Pol oscillator, as $\mu$ gets smaller the limit cycle takes the shape of a circle and  for large values of $\mu$ the shape of the limit cycle appears more rectangular.
	Given the rectangular shape of the limit cycle of the hybrid model in \eqref{eq:Temperature}-\eqref{eq:switch}, we chose a large value of $\mu=100$ such that the limit cycle of  \eqref{eq:Continuous_mdl_v1}  is similar to the one in Fig. \ref{fig:Fig1_Temp_switching}c while oscillating approximately at the same frequency as \eqref{eq:Temperature}.
	
	Once the  value of $\mu$ is fixed, the small difference in the deadband that translates into frequency deviation can be compensated for by the parameter $\s <\delta$.
	The parameter $\s$ can quickly be determined through numerical simulations as follows.
	Knowing that $\s \in [0,\delta)$, one can sample $n$ values of $\s$ over this interval, simulate the dynamics \eqref{eq:Continuous_mdl_v1} for each value of $\s$, and compare the oscillation frequencies at each value to the nominal frequency of the hybrid state model in \eqref{eq:Temperature} to determine the most appropriate value of $\s$. This is only done once using the TCL with a natural frequency corresponding to the average frequency of the population of TCLs considered. For the parameters in the Table 1, $\s = 0.1454$.
	
\begin{figure}[!h]
		\centering
		\includegraphics[width=1\linewidth]{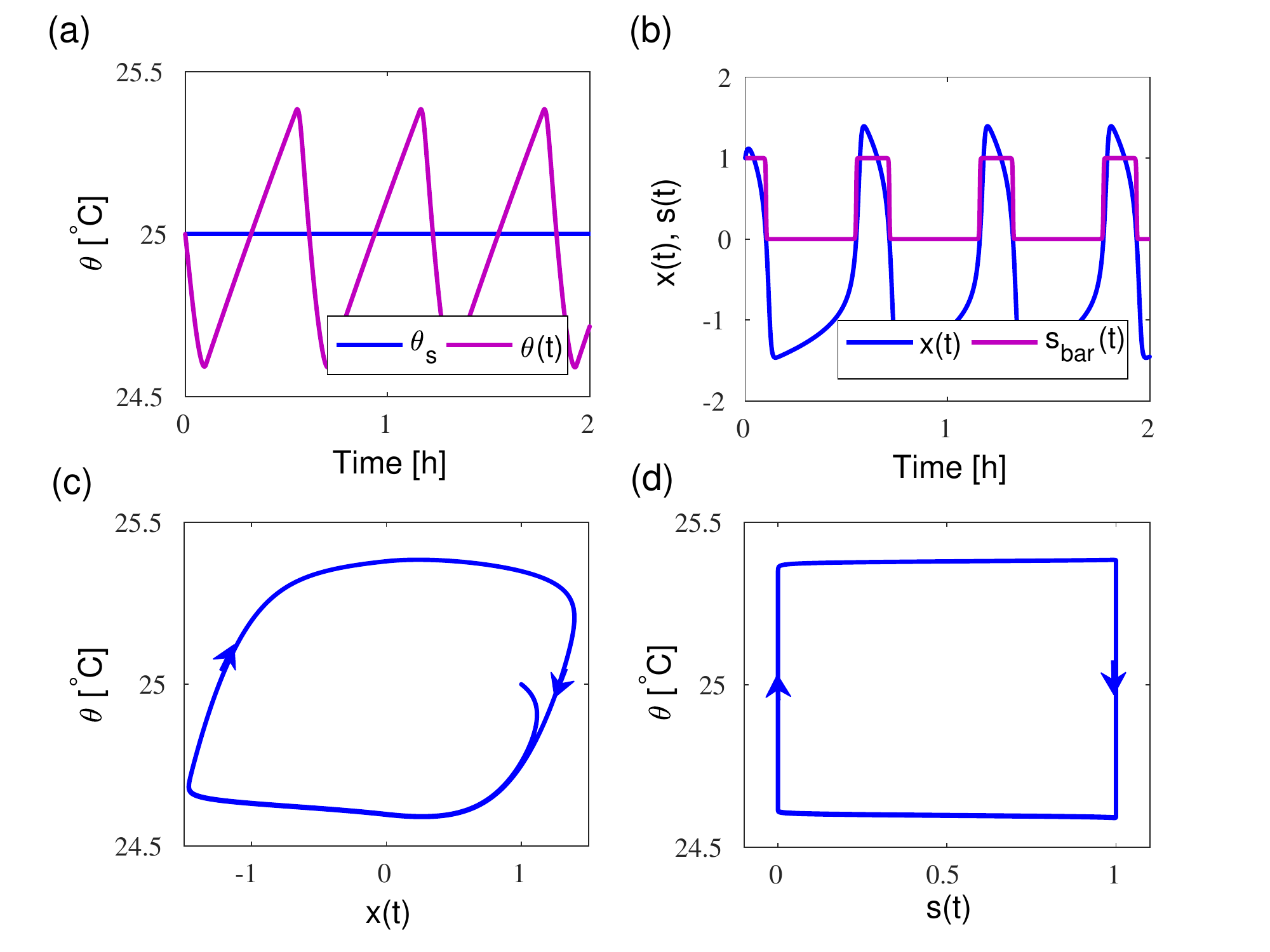}
		\caption{Simulation of the system in \eqref{eq:Continuous_mdl_v1}. (a) Evolution of temperature $\t(t)$ around the setpoint $\t_s = 25^{\circ}$C. (b) The switching state variable $x(t)$, and the switching function $\bar{s}(t)$. (c, d) Limit cycles for both $\t (t)$ {\em vs.} $x(t)$ in (c) and $\t(t)$ {\em vs.} $\bar{s}(t)$ for the new switching function \eqref{eq:switch_fct} in (d). The parameters used are in Table 1, but with $C = 2$ kWh$/^{\circ}$C and $\delta = 1.0^{\circ}$C. }
		\label{fig:Fig3_Plt_mdl_v2}
		\vspace{-1ex}
	\end{figure}

	In order to obtain a continuous switching function $\bar{s}(t)$ from the variable $x(t)$ that is as similar as possible to the switching function $s(t)$ in the hybrid model,  we use an approximation of the Heaviside step function.
	Such step function approximations have been used in a similar way in the Morris-Lecar neuron model, whose behavior is similar to the Hodgkin-Huxley spiking neuron model \cite{Ermentrout96}.  The resulting switching function approximation is given by
	\begin{equation}\label{eq:switch_fct}
	\bar{s}(t) = \frac{1}{2}(1+\tanh(k x)),
	\end{equation}
	where the parameter $k \in [5,10]$ controls the sharpness of the switching action. As shown in Fig. \ref{fig:Fig3_Plt_mdl_v2}d, the phase portrait is similar to that of the hybrid-state model shown in Fig. \ref{fig:Fig1_Temp_switching}c.
	
\begin{figure}[!h]
		\centering
		\includegraphics[width=1\linewidth]{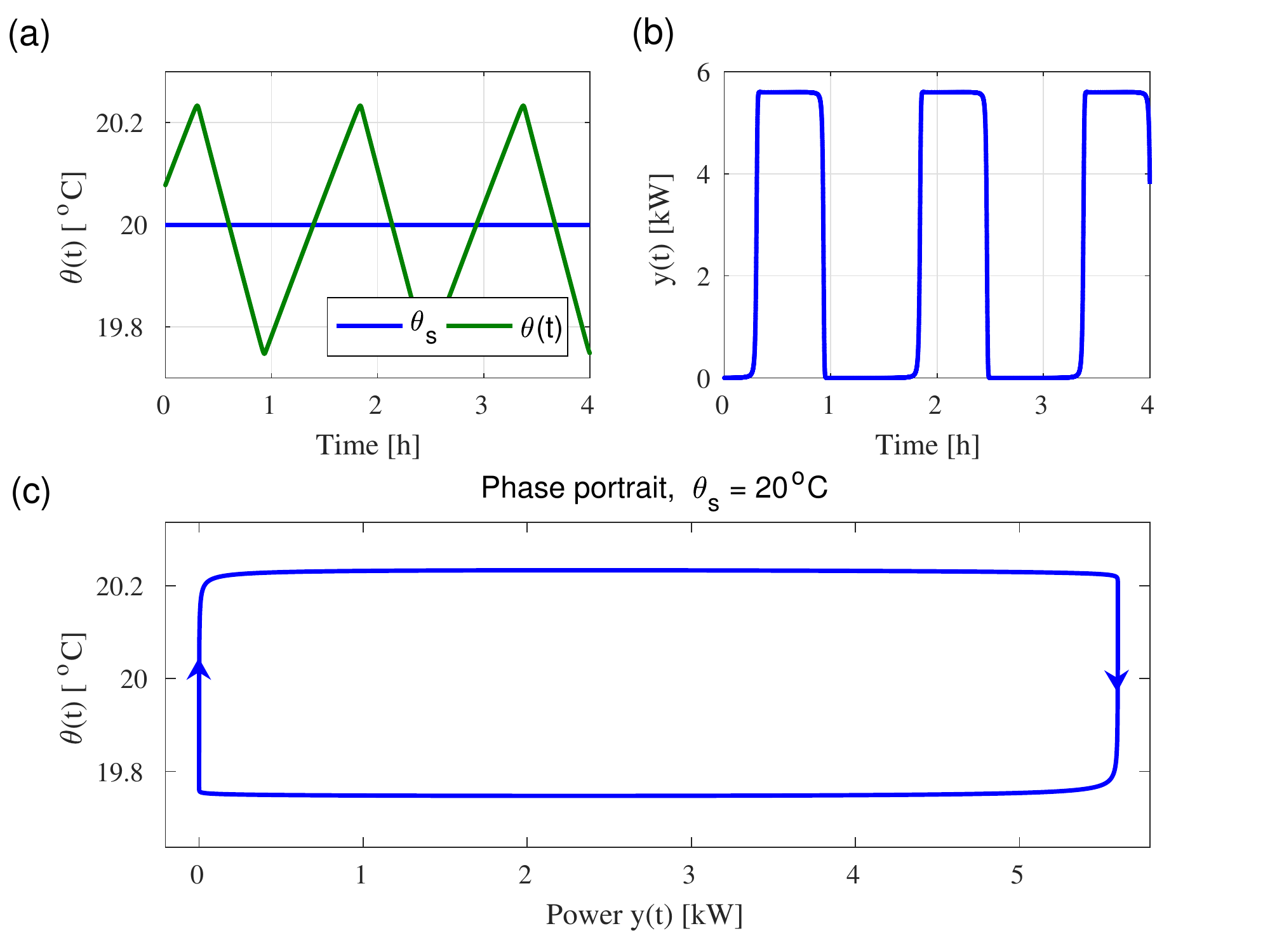}
		\caption{Simulation of the system in \eqref{eq:Continuous_mdl_v2}. (a) Evolution of the temperature $\theta(t)$ around the setpoint $\theta_s=20^{\circ}$C. (b) Evolution of the electric power $y(t)$ drawn by the TCL. (c) TCL phase portrait. The parameters used are given in Table 1.}
		\label{fig:Fig4_Plt_mdl_power}
		\vspace{-1ex}
	\end{figure}

	The model given in \eqref{eq:Continuous_mdl_v1} reproduces the dynamical behavior of the hybrid-state model in \eqref{eq:Temperature}.
	However, our goal is to design an ensemble control of a TCL population such that the aggregate power closely tracks a given reference power.
	Therefore, we modify the model in \eqref{eq:Continuous_mdl_v1} by substituting for the state variable $x(t)$ with the instantaneous power $y(t)$. The derivation of this model is provided in Appendix \ref{sec:Appendix_A}.  The resulting model is given by
	\begin{equation}\label{eq:Continuous_mdl_v2}
	\begin{split}
	\dot{y}(t) & = \mu k \left( (\frac{\delta}{2}+\s) \bar{y} - \frac{1}{3}\bar{y}^3 +\t - \t_s \right)(1-\frac{\eta}{P}y)y, \\
	\dot{\t}(t) & = -\frac{1}{CR} \left(\t - \t_a +\eta y(t)R \right),
	\end{split}	
	\end{equation}
	where $\bar{y}(t)  = -\frac{1}{k}\ln(\frac{P}{\eta y}-1)$.
	Note that the nominal TCL power in Table 1 is the average energy transfer rate and not the electric power.
	Hence, the actual electric power consumed by a TCL when running at full load is given by the average energy transfer rate $P$ divided  by the coefficient of performance $\eta$.
	The aggregate electric power of an ensemble of $N$ TCLs described by \eqref{eq:Continuous_mdl_v2} is given by
	\begin{equation}\label{eq:P_aggr_v2}
	P_{agg}(t) = \sum_{i=1}^N y_i(t).
	\end{equation}
	The evolution of the TCL's internal temperature in \eqref{eq:Continuous_mdl_v2}, its instantaneous power and the limit cycle are shown in Fig. \ref{fig:Fig4_Plt_mdl_power}.
	
	\subsection{TCL Phase Model}
	\label{sec:TCL_Phase_mdl}
	
\begin{figure}[!h]
		\centering
		\includegraphics[width=1\linewidth]{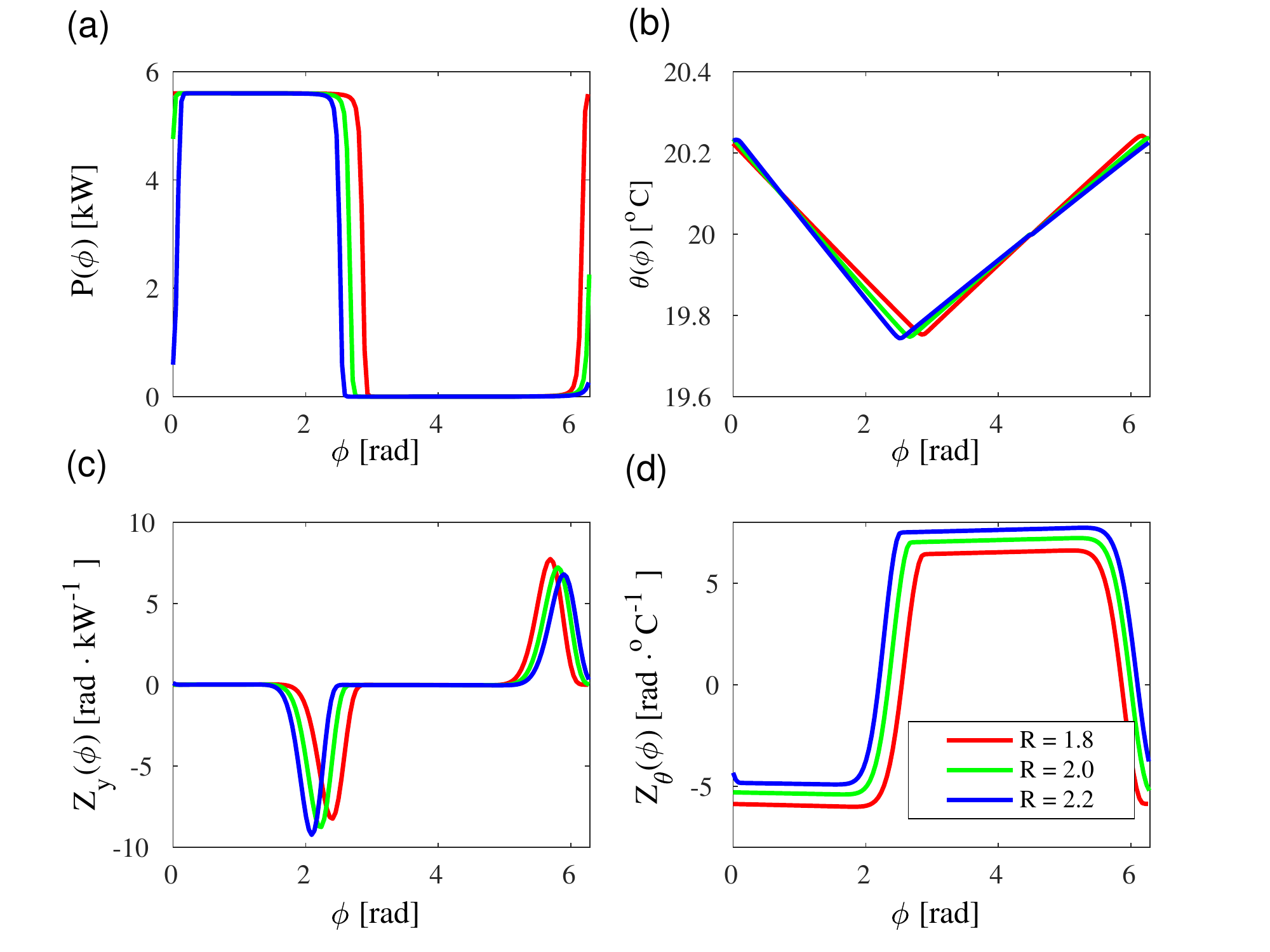}
		\caption{Phase response curves of the power and temperature for a range of thermal resistance $R \in [1.8, 2.2]$. (a) and (b) show the variation of the power and temperature as a function of the phase $\phi$. (c) and (d) show their phase sensitivity functions $Z_y(\phi)$ and $Z_\theta(\phi)$, respectively, for the dynamics in \eqref{eq:Continuous_mdl_v2}. The parameters used are shown in Table 1. }
		\label{fig:PRC_p_V2}
		\vspace{-1ex}
	\end{figure}

	In Section \ref{sec:Basic_D_mdl}, we have shown that the unforced cycling dynamics of a TCL can be represented by a phase model $\dot{\phi} = \omega$, where $\omega$ is the natural frequency. However, in the presence of an external control input $u(t)$, the phase model takes the form
	\begin{equation}\label{eq:phase_mdl1}
	\dfrac{d\phi}{dt}(t) =  \omega + \epsilon Z(\phi)u(t),
	\end{equation}
	where  $Z(\phi)$ is the phase sensitivity function, also known as phase response curve (PRC) and $\epsilon$ is the intensity of the perturbation or control input $u(t)$ \cite{Nakao15}.  For  \eqref{eq:Continuous_mdl_v1} and \eqref{eq:Continuous_mdl_v2}, the PRCs are vectors  $Z(\phi) =  (Z_s(\phi),  Z_\t(\phi))$  and $Z(\phi) =  (Z_y(\phi),  Z_\t(\phi))$, respectively, where $Z_s(\phi),\ Z_y(\phi)$ and $Z_\t(\phi)$ are  phase sensitivity functions of the switching variable $x(t)$, the instantaneous power $y(t)$, and the temperature $\t(t)$, respectively (see Fig. \ref{fig:PRC_p_V2}).  Because we are only interested in controlling one state variable of the TCL e.g., the switching $s(t)$ or the power $P(t)$, the PRC in \eqref{eq:phase_mdl1} will be taken as one of the scalar functions $Z(\phi) = Z_s(t)$ or $Z(\phi) = Z_y(t)$, and the scalar control input will be a temperature signal to offset the set-point. If on the other hand we were to consider the phase model of the temperature $\t(t)$, the PRC will be $Z(\phi) = Z_\t(t)$ and the corresponding control will be the input power.

	The natural frequency is computed as $\omega = 2 \pi/T$ using the period $T$ in \eqref{eq:T} and it is the same for all three variables $s(t),\ P(t)$ and $\t(t)$.
	The PRC itself must be computed numerically, using for example the method of the adjoint which requires the computation of the Jacobian of the dynamics of the system described by a continuous function. This method consists of linearizing the system around its periodic orbit $\Gamma(t)$ and solving the adjoint equations with a backward integration \cite{Nakao15, Ermentrout96}. A standard software package used for the computation of the PRC is the XPPAUT \cite{ermentrout2002simulating}.
	The ability to compute the PRC using the method of the adjoint was one of the reasons for the derivation of the continuous models described in Section \ref{sec:2D_mdl}.
	Another important reason for such modeling is the ability to access the switching variable $x(t)$ in \eqref{eq:Continuous_mdl_v1} or the power variable $y(t)$ in \eqref{eq:Continuous_mdl_v2} in order to characterize their dependence on a stimulus $u(t)$. The phase sensitivity quantifies how much the phase of these variables changes in response to impulse $I(t)$.
	In Appendix \ref{sec:Appendix_B}, we review some basics of the phase reduction theory as well as different techniques for determining the phase sensitivity curve of an oscillatory system.
	
	From the phase model \eqref{eq:phase_mdl1}, one can see that the control action is applied through the PRC, consequently the oscillator phase will either be advanced or delayed.
	This means that the time at which the TCL turns ON or OFF can either be delayed or advanced by $\Delta T$  when an appropriate control input is applied. The implication is that the duty cycle of the input power to the TCL can be modulated, and therefore its instantaneous or average power can be increased or decreased accordingly.
	
	The phase sensitivity functions for the temperature and power evolution described by \eqref{eq:Continuous_mdl_v2} are shown in Fig. \ref{fig:PRC_p_V2} for various values of the thermal resistance $R$.
	This shows that the parameter heterogeneity of the TCL ensembles will cause each unit to respond differently to a common control.
	The PRC of \eqref{eq:Continuous_mdl_v1} looks similar to that of \eqref{eq:Continuous_mdl_v2} except for the amplitude of $Z_s(\phi)$ which is smaller than $Z_y(\phi)$.
	
	\section{ Simulation Comparison of TCL Models}
	\label{sec:Sim_mdl}
	
\begin{figure}[!h]
		\centering
		\includegraphics[width=1\linewidth]{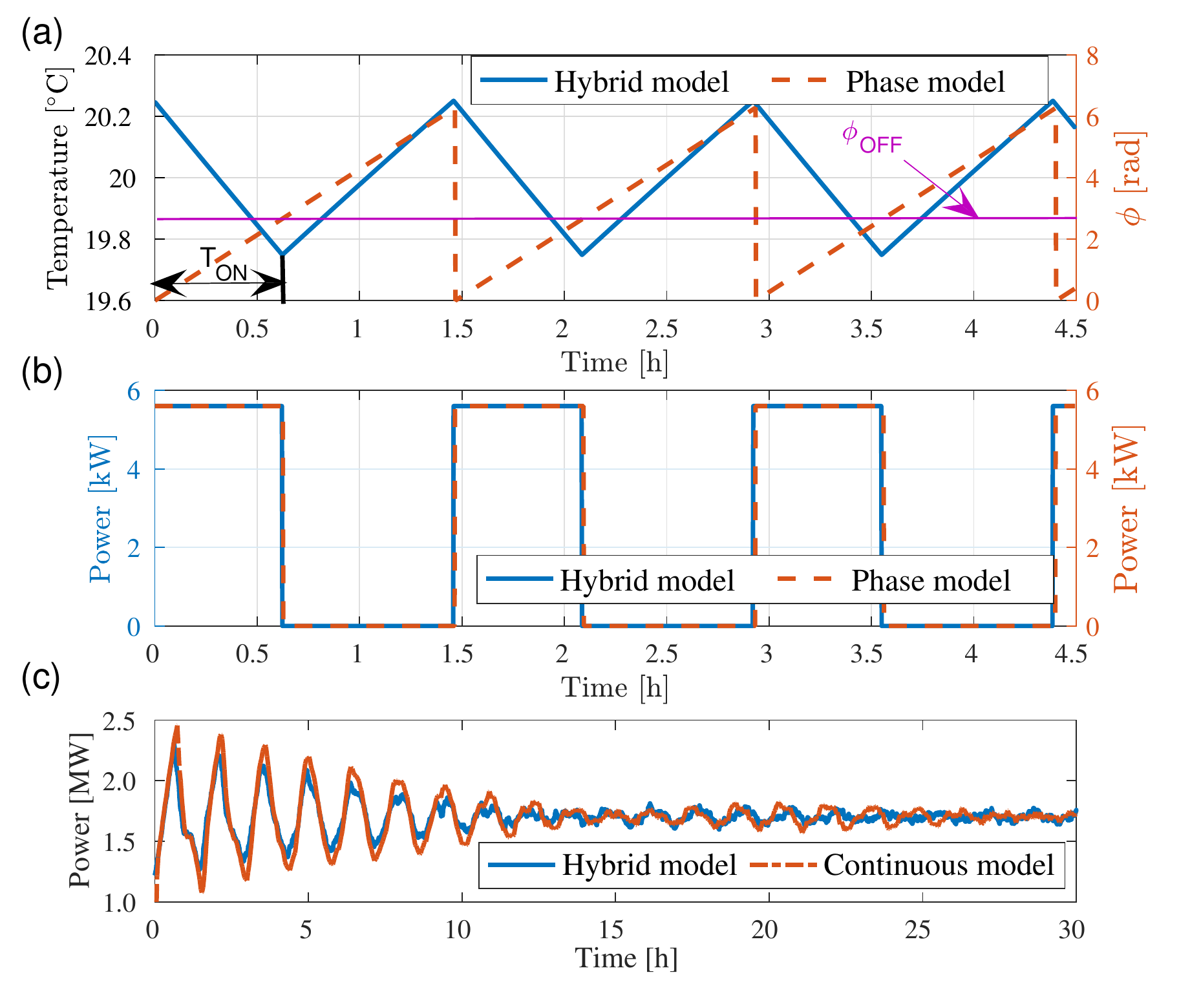}
		\caption{Comparison between the Hybrid model and its derived phase model representation.
			(a) Time evolution of the temperature (blue, solid) and the phase evolution (red, dashed). (b) Time evolution of the power consumed by the TCL for both models. Note that the TCL turns OFF exactly after being ON for time $t=T_{ON} $ and the switching phase is $\phi_{OFF}$.
			(c) Aggregate power of 1000 heterogeneous TCLs described by the hybrid and continuous models \eqref{eq:Temperature} and \eqref{eq:Continuous_mdl_v2}, respectively.
			Simulated with the  parameters in Table 1. }
		\label{fig:Hmdl_Phasemdl_comp}
		\vspace{-1ex}
	\end{figure}

	In this section, we provide simulation results that are intended to show how well the continuous model \eqref{eq:Continuous_mdl_v2} proposed in Section \ref{sec:2D_mdl} approximates the hybrid model dynamics in \eqref{eq:Temperature}-\eqref{eq:switch}.
	We first compare the phase model in \eqref{eq:phase_mdl1} to the hybrid-state model in \eqref{eq:Temperature}-\eqref{eq:switch} by plotting the time evolutions of the phase $\phi$ and the temperature together in Fig. \ref{fig:Hmdl_Phasemdl_comp}a. The phase $\phi_{OFF}$ indicates where the TCL turns OFF after being ON for a time $T_{ON}$. The  input powers for both models are shown Fig. \ref{fig:Hmdl_Phasemdl_comp}b, with the control $u(t)=0$.
	
	Second, using the data in Table 1, we simulate the aggregate power of 1,000 heterogeneous TCLs for a period of 30 hours. The initial values of temperature and the parameters $C$, $R$ and $P$ were randomly distributed uniformly within $\pm5\%$ of their nominal values. The results in Fig. \ref{fig:Hmdl_Phasemdl_comp}c show good agreement between the transient oscillations of both the hybrid and continuous models, and the stationary variation about the long-run mean which occurs after 10 hours appears similar as well.

	Now that we have shown that the phase model \eqref{eq:phase_mdl1} captures the cycling dynamics of a TCL with sufficient accuracy, in the next section we will analyze the synchronization properties of phase model representations of TCLs, and then derive a PRC-based control policy.
	
	\section{PRC-Based Control Policy}
	\label{sec:ControlPolicy}
	
	Direct control of TCLs by the Balancing Authority can enable regulation of power consumed by such loads on a distribution subsystem over faster time scales than the price response approach allows.  Among possible control architectures proposed in the literature, one may find centralized, hierarchical, and distributed controllers  \cite{Callaway11}.
	Implementing a centralized feedback control policy to modulate the power consumption of an ensemble of TCLs is challenging. It requires the measurement and transmission of the states of each TCL to a centralized controller, resulting in extra cost for measurement equipment and communication channels that need to be established, which has also raised some privacy concerns \cite{sinitsyn2013safe}.
	Other issues such as temporary synchronization arise when a step change in the control input $u(t)$ is applied to a large number of TCLs. This causes the aggregate power to strongly oscillate for up to a few hours before settling to the desired steady state value \cite{ mehta2014safe, Perfumo12, Kundu11}.
	In this section we analyze and implement a PRC-based control policy that effectively regulates the aggregate power of an ensemble of TCLs on a relatively fast time scale while preventing undesirable power fluctuation that can be caused by temporary synchronization.
	
	\subsection{Control of a Single TCL}  \label{sec:SingleTCLControl}
	
	Phase reduction theory has been used in various scientific areas including physics, neural engineering and biology. This powerful technique has enabled the analysis of synchronization properties of limit cycle oscillators \cite{Nakao15}. One can find various applications in the literature such as entrainment of chemical oscillator \cite{harada2010optimal}, optimal entrainment of neural oscillator ensembles \cite{Zlotnik12} and phase advance or delay in circadian oscillators where light is used as a control input \cite{efimov2009controlling}. Most applications of the phase reduction theory consist of either controlling the phase or the frequency of oscillators e.g., controlling the spiking time of neurons \cite{dasanayake2011optimal}, and different control techniques  have been developed for that purpose \cite{Zlotnik12, efimov2009controlling}.
	
\begin{figure}[!h]
		\centering
		\includegraphics[width=1\linewidth]{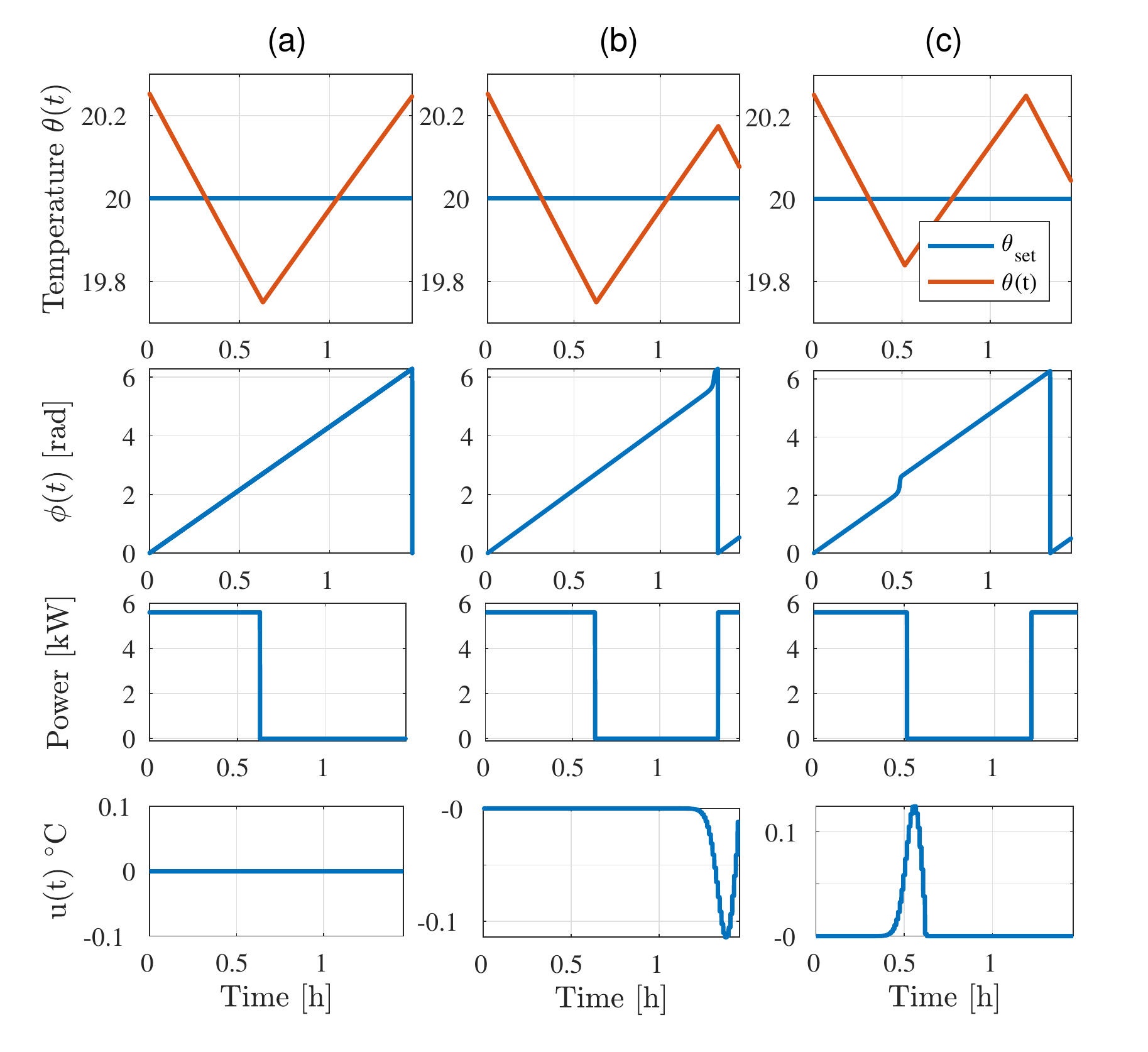}
		\caption{Illustration of the phase advances induced by the control input $u(t)$. Column (a) shows the unperturbed system. Columns (b) and (c) show the controlled systems. In (b), the control switches the TCL ON before its normal turn ON time by temporarily decreasing the set-point. The opposite happens in (c). The first and second rows of plots show the temperatures and their corresponding phases. The third and fourth rows show the corresponding power and control waveforms.}
		\label{fig:Fig_IllustrationOfPowerModulation}
		\vspace{-1ex}
	\end{figure}

	In the present application, in addition to regulating the switching frequency of a TCL, we wish  to also modulate the duty cycle of the input power such that the average power consumption during a period of time $T$ is either increased or decreased based on a regulation signal $\xi(t)$. By appropriately switching the TCLs ON and OFF as illustrated in Fig. \ref{fig:Fig_IllustrationOfPowerModulation}, one can modulate the aggregate power of an ensemble over a short time scale without impacting the average temperature of the individual units over the long run.  The controls used to produce the results in Fig. \ref{fig:Fig_IllustrationOfPowerModulation}  are of the froms $u(t) = Z_{+}(\phi)\xi(t)$ or $u(t) = -Z_{-}(\phi)\xi(t)$ for increasing or decreasing the power consumption, respectively, where $Z_{\pm}(\phi)$ represent the positive and negative parts of the switching PRC ($Z_s(\phi)$), and the regulation signal $\xi(t) = \mp 0.1^{\circ}C$ was used to increase and decrease the power by forcing the TCL to switch either ON (Fig. \ref{fig:Fig_IllustrationOfPowerModulation}b) or OFF (Fig. \ref{fig:Fig_IllustrationOfPowerModulation}c) early.
	
\begin{figure}[!h]
		\centering
		\includegraphics[width=1\linewidth]{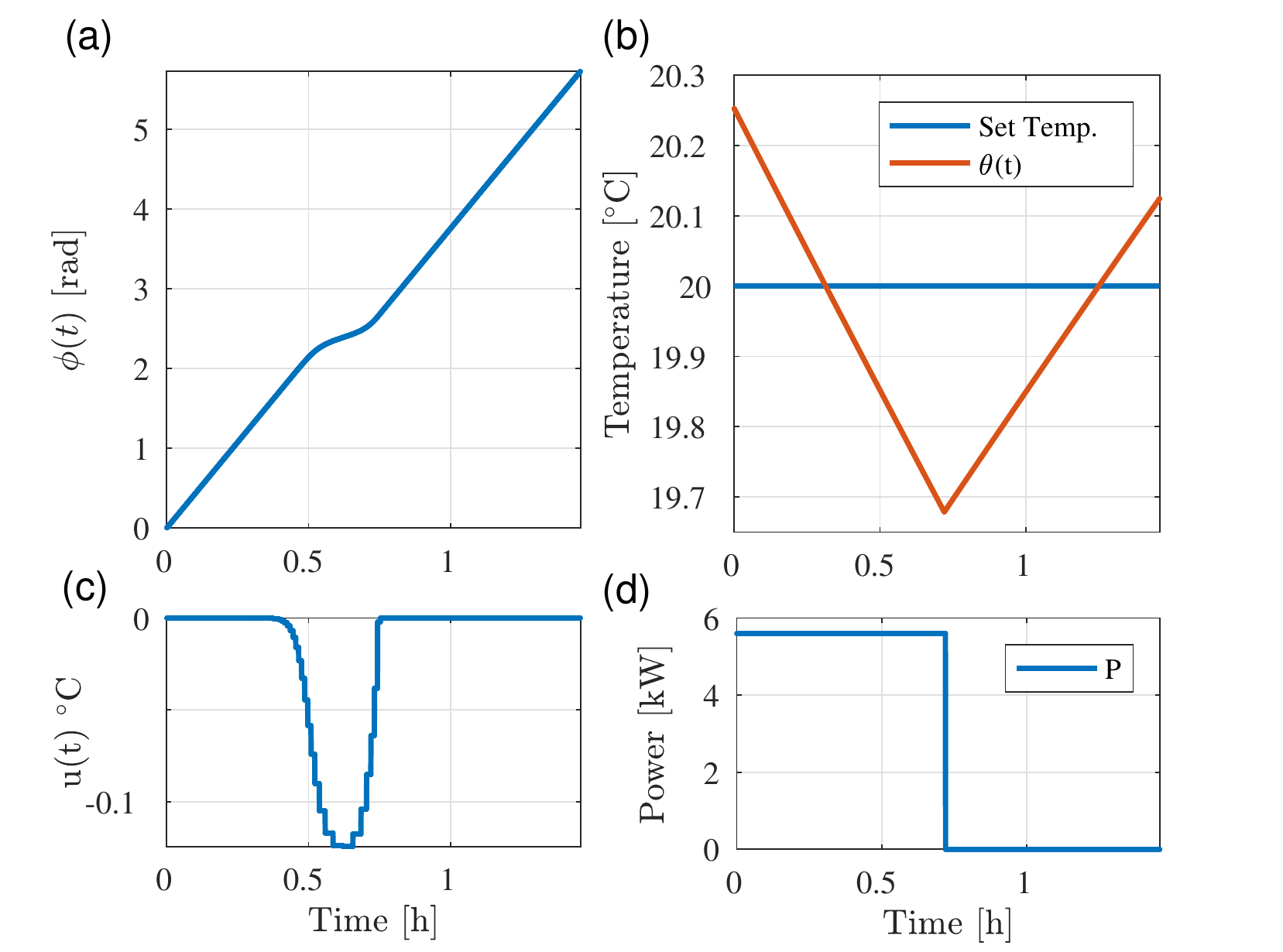}
		\caption{Illustration of a phase delay induced by the control $u(t)=-Z_{-}\xi(t)$, with $\xi(t)=-0.1^{\circ}C$. (a) Evolution of the phase $\phi(t)$ showing a slow down (delay) when the control is applied. (b) Evolution of the temperature showing that the TCL stays ON longer, hence the temperature went below the lower limit of $19.75^{\circ}$C. (c) and (d) show the control input (shift of the set-point temperature) and the TCL power over that period of time, respectively.}
		\label{fig:Fig_PhaseDelayExample}
		\vspace{-0ex}
	\end{figure}

	The phase advances induced by $\xi(t)$ in the phase model \eqref{eq:phase_mdl1} are depicted in the second row of Fig. \ref{fig:Fig_IllustrationOfPowerModulation}.
	By switching a large number of TCLs ON, the aggregate power given by \eqref{eq:P_aggr} will instantaneously increase, and conversely switching them OFF will decrease the aggregate power. \
	It is equally possible to delay the phase which results in the TCL staying ON or OFF longer than it would naturally. An example is illustrated in Fig. \ref{fig:Fig_PhaseDelayExample}.	
	
	\subsection{Control of an Ensemble of TCLs}
	\label{sec:EnsembleTCLControl}
	
	One of the main challenges encountered when designing an open-loop control policy for a large population of TCLs is the temporary synchronization that can be caused by a sudden change of the set-point. Although the changes can be small (0.1-0.5$^{\circ}$C) and barely noticeable by the customer, they can induce large power fluctuations \cite{Callaway09, zhang2013aggregated}.  Alternatively, such policies can track the aggregate power reference closely when it is relatively slowly varying and smooth.  The thermostat in this case must be assumed to be adjustable with infinitesimally fine precision, although this requirement may be relaxed in application. To address the synchronization issues, various solutions have been proposed. For instance, a feedback control for TCLs with finite thermostat resolution was explored in \cite{Perfumo12}.
	Among the open-loop control policies, \cite{totu2014demand} developed a switching-fraction broadcast signal that determines the number of loads to be switched ON or OFF by solving an optimization problem with a multilinear objective.

\begin{figure}[!h]
		\centering
		\includegraphics[trim=140 200 140 200,clip,width=1\linewidth]{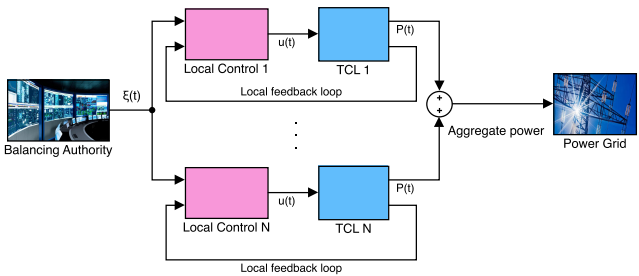}
		\caption{Block diagram of the control architecture. The BA regulation signal $\xi(t)$ is sent to all the TCLs in the population. This signal requests that each TCLs changes its power consumption by a given fraction that, if aggregated, will compensate for demand on the power grid. The local controller receives $\xi(t)$ as well as the states of the TCL, then determines the appropriate control input $u(t)$.}
		\label{fig:ControlBlockDiagram}
		\vspace{-1ex}
	\end{figure}

	The open-loop control policy proposed herein addresses a similar issue as encountered in the application of the safe methods of \cite{sinitsyn2013safe, mehta2014safe}  or the priority stack framework \cite{Hao15, Vrettos2012load}. Specifically, we aim to prevent excessive aggregate power fluctuations caused by synchronization of a large number of TCLs in the ensemble that in turn arise from sudden changes of the control signal (e.g., set-point temperature).
	The synchronization is prevented by letting the phase sensitivity function dictate the response of a unit to a quickly changing control signal.  Note that because of TCL heterogeneity, the PRCs of all the units in a population are different.   The PRC-based control architecture is presented in Fig. \ref{fig:ControlBlockDiagram}. It is assumed that each TCL can measure its state variables (temperature and switching status) and has knowledge of its own phase sensitivity function $Z(\phi)$. Hence the controller can generate the corresponding control $u(t)$ in response to a global regulation signal $\xi(t)$ from the BA. 
	
	The open-loop control structure illustrated in Fig. \ref{fig:ControlBlockDiagram} presents several practical advantages. First of all, it considerably limits the information complexity required at the Balancing Authority  level in the sense that no feedback is required to form the reference signal $\xi(t)$. This also limits the computation and communication costs that would have occurred if feedback was needed. By suppressing the need for a two-way communication, this control policy also addresses privacy concerns \cite{sinitsyn2013safe}.  The assumptions that we make on capabilities of the TCL and communication with the BA can be summarized as 1) the BA  has knowledge of the power utilization of the TCL ensemble and its capacity to service demand response, 2) in some instances the BA can totally or partially estimate the aggregate power of the ensemble being controlled, 3) each TCL has knowledge of its own PRC and can measure its internal states, and 4) each TCL is equipped with a control unit and a thermostat that has fine resolution on the deadband.
	
	As a proof of concept, we propose a PRC-based integral controller, and evaluate its performance against a traditional integral controller and the direct control. By direct control, we refer to situations with feedback and no local controller, in which the signal $\xi(t)$ directly control the TCL ensemble. We further show its efficacy by tracking a real Area Control Error (ACE) signal taken from the Bonneville Power Administration (BPA) website \cite{BPA}.
	The proposed controller is of the form
	\begin{equation}\label{eq:control_diff_form}
	\dot{u}(t) = I_1 [I_2 \text{sgn}(\tfrac{1}{2}-s(t)) Z(\phi(t)) \xi(t) - u(t)],
	\end{equation}
	where {\em{sgn}} is the signum function that extracts the sign of $1/2-s(t)$, $I_1$ and $I_2$ are control gains. Discretizing the differential form of the control in \eqref{eq:control_diff_form}, we arrive at $u(t)=u_k$ for $t\in[t_k,t_{k+1}]$ where
	\begin{equation}\label{eq:control}
	u_{k+1} = u_k+I_1 h [ I_2 \text{sgn}(\tfrac{1}{2}-s_k) Z(\phi_k)  \xi_k - u_k ],
	\end{equation}
	where $h = t_{k+1}-t_k$ is the time step, and we have substituted dependence on the discrete time $t_k$ by the subscript $k$ for simplicity. The control equation \eqref{eq:control} describes what each local controller in Fig. \ref{fig:ControlBlockDiagram} is doing when the regulation signal $\xi(t)$ is received.
	To understand how this control policy is able to track a regulation signal without excessive synchronization of TCL dynamics, observe that $\xi(t)$ enters the control signal through a product with the PRC ($Z(\phi) = Z_y(\phi)$), which is different from one TCL to another (see Fig. \ref{fig:PRC_p_V2}).  This implies that each TCL will respond differently according to its own parameters. Without loss of generality, we may assume that at time $t=t_k$, $u_k = 0$ and $\xi_k > 0$, requesting that the TCLs reduce their power consumptions by increasing their set-points by a small fraction of the nominal value. Given the shape of each PRC, the requested change will not be instantaneous for all TCLs.  Depending on where the TCLs are in their cycles, some will switch their status right away while others will do so with a delay that is function of the PRC. The gains $I_1$ and $I_2$ control the response time of the controller and the steady state error, respectively.
	The choice of an integral controller of the form \eqref{eq:control} was motivated by the desire to minimize higher order harmonics present in the PRCs.

	
	\subsection{Analysis of Temporary Synchronization in TCL Ensembles }
	\label{sec:Analysis}
	The potential demand response (DR) service that an ensemble of TCLs could be used to provide to a power grid is limited by many factors such as the specified limits for maintaining customer quality of service, the number $N$ of TCLs in the population  (the power capacity of a TCL ensemble  increases with $N$), and crucially the frequency bandwidth in which DR can be extracted.  Temporary synchronization that causes undesirable fluctuation of the aggregate power is a consequence of the bandwidth limit.
	
\begin{figure}[!h]
		\centering
		\includegraphics[width=1\linewidth]{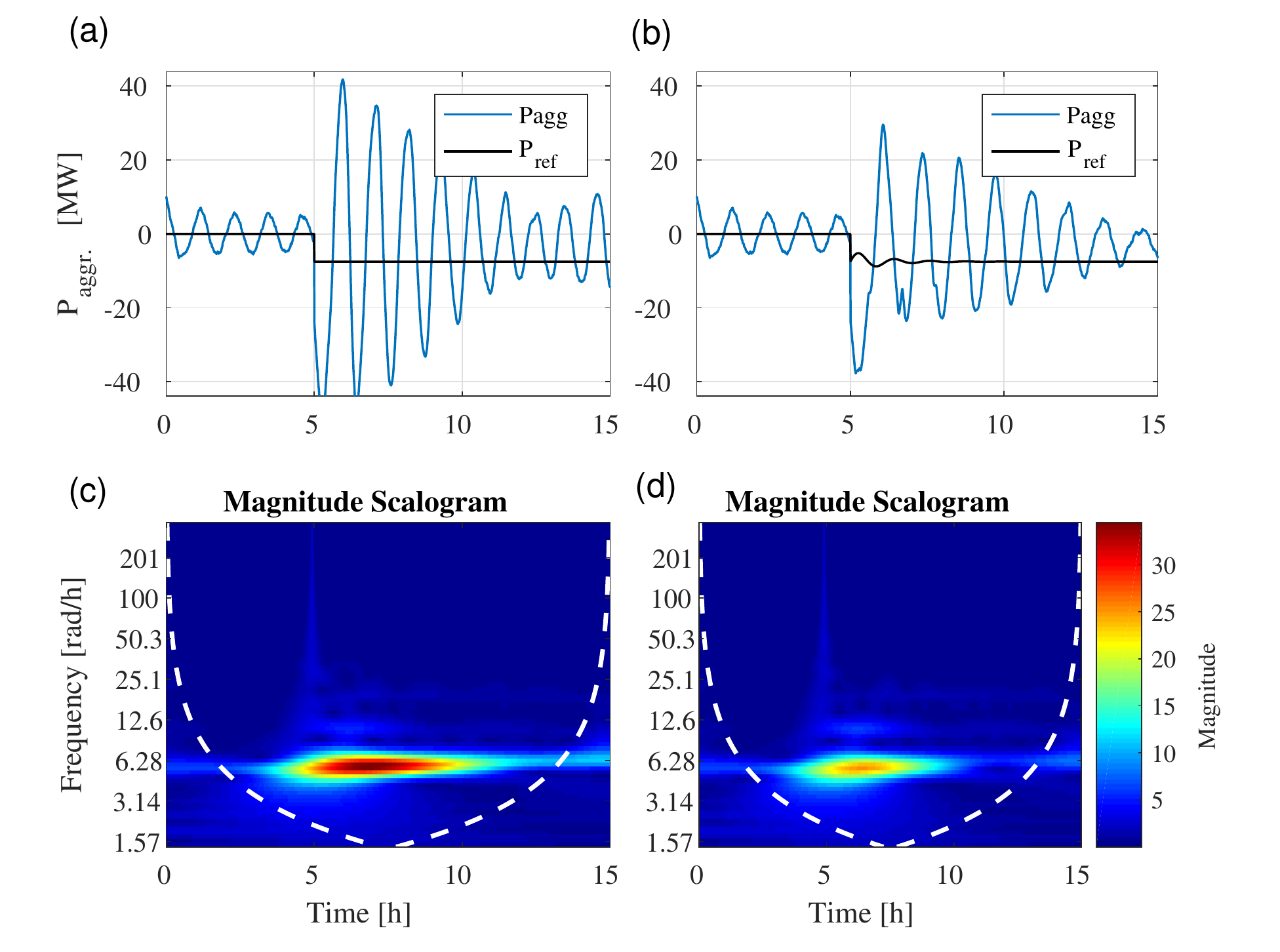}
		\caption{Response of a heterogeneous TCL ensemble to a step change in the reference power. (a)-(c) Aggregate power response to a step change and its wavelet transform magnitude scalogram. (b)-(d)  Aggregate power response to a step change with an opposing decaying sinusoidal of the same frequency as the induced oscillations. Comparing (c) and (d) reveals that the introduction of a decaying sinusoid with sign opposite the induced oscillation reduces the power content of the undesired oscillations considerably.
		}
		\label{Fig_wavelet_Power_stepChange}
		\vspace{-1ex}
	\end{figure}

	In this section, we provide some useful tools that elucidate the synchronization behavior of TCL ensembles. Observe what happens when a population of TCLs is made to track a step change in the reference power (Fig. \ref{Fig_wavelet_Power_stepChange}). Before the step change is applied at time $t = 5h$, the initial conditions and the finite number of TCLs in the ensemble cause the aggregate power to oscillate with a frequency $\omega_0 \approx  1/N \sum_{i=1}^{N} \omega_i  $, where the $ \omega_i$'s are the natural frequencies of the TCLs in the population. Using the wavelet transform, we compute the power spectrum of the aggregate power as a function of time. It appears that the TCL ensemble naturally has damped oscillations with a mean frequency $\omega_0=5.69$ rad/h and that a step change at $t=5h$ amplifies these oscillations (Fig. \ref{Fig_wavelet_Power_stepChange}c). Observe that by superimposing a decaying sinusoidal signal whose sign is opposed to the step change, the power or amplitude of the oscillations greatly decreases (Fig. \ref{Fig_wavelet_Power_stepChange}b-d), which implies that it is possible to design a control that can change the power usage of the ensemble over a short time scale while minimizing the unwanted oscillations.
	
\begin{figure}[!h]
		\centering
		\hspace{-20pt}
		\includegraphics[width=1\linewidth]{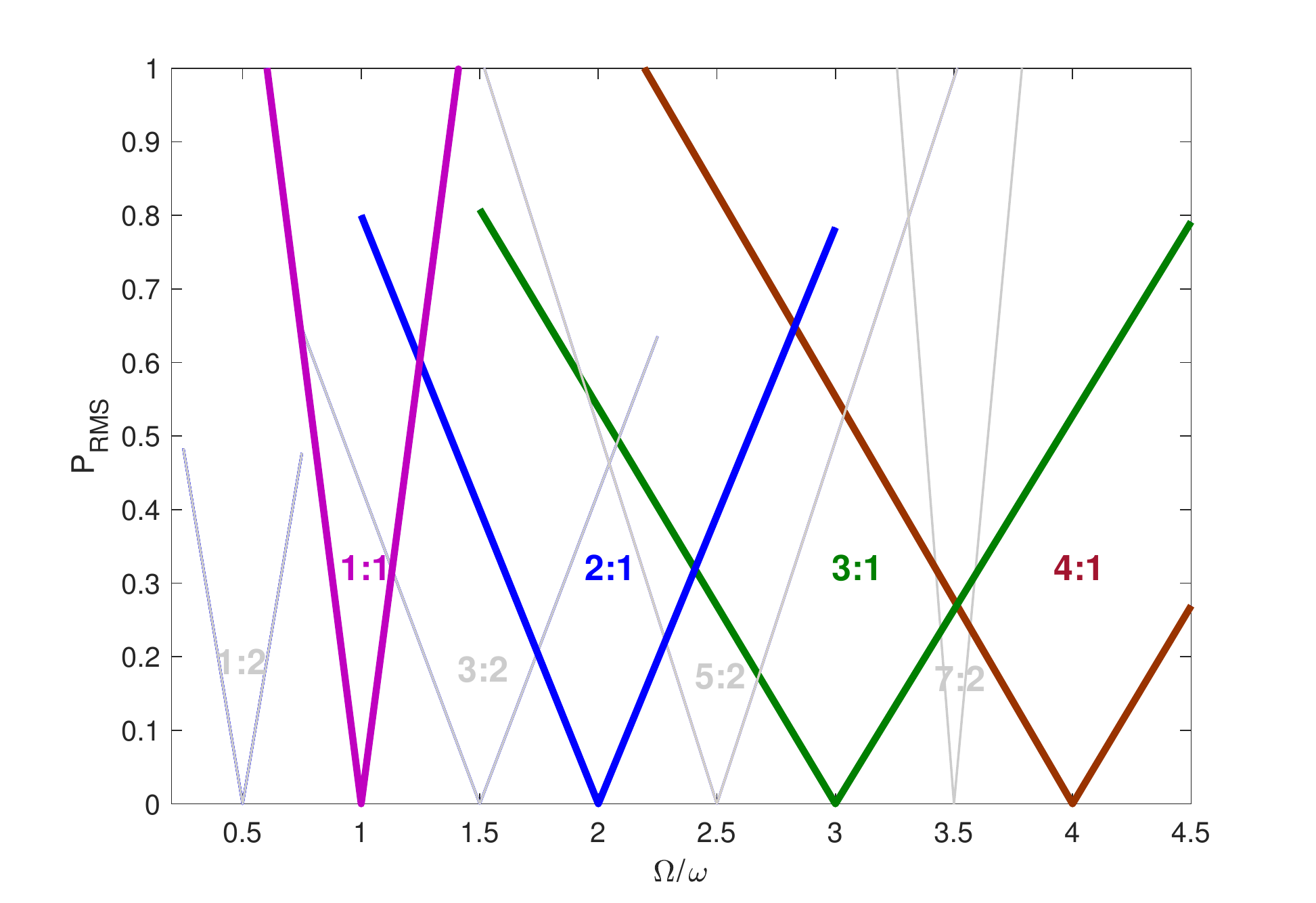}
		\caption{Theoretical Arnold tongues for a sinusoidal input with frequency $\Omega$ applied to one TCL with natural frequency $\w$. The entrainment ratios are indicated as N:M, with $\Omega = N/M\omega$. For this example, the TCL natural frequency is $\omega= 7.95$ rad/h,  the parameters  in Table 1  were used with different thermal capacitance $C=1.8$ kWh/$^{\circ}$C and the deadband $\delta=1.5^{\circ}$C, respectively.
		}
		\label{Fig_TheoreticalArnold_SinInput}
		\vspace{-1ex}
	\end{figure}

	It is crucial to note that the step at $t=5h$ behaves like an impulse stimulus whose power content extends over all the frequencies and the population is strongly excited by the frequency closer to its natural mean frequency.  It appears as if all the TCLs are now oscillating with the same frequency $\omega_0$. This phenomenon is known in the study of rhythmic systems  as frequency entrainment \cite{harada2010optimal,zlotnik2016phase,tanaka2014optimal}. The phase model representation of nonlinear oscillators becomes highly valuable in this case because the phase sensitivity function $Z(\phi)$ can be used to provide the theoretical linearized limits of the entrainment region commonly referred to as Arnold tongue (see Fig. \ref{Fig_TheoreticalArnold_SinInput}) \cite{shirasaka2017phase, granada2009phase, tanaka2015optimal}. For more details see Appendix \ref{sec:Appendix_C}.

\begin{figure}[!h]
		\centering
		\includegraphics[width=1\linewidth]{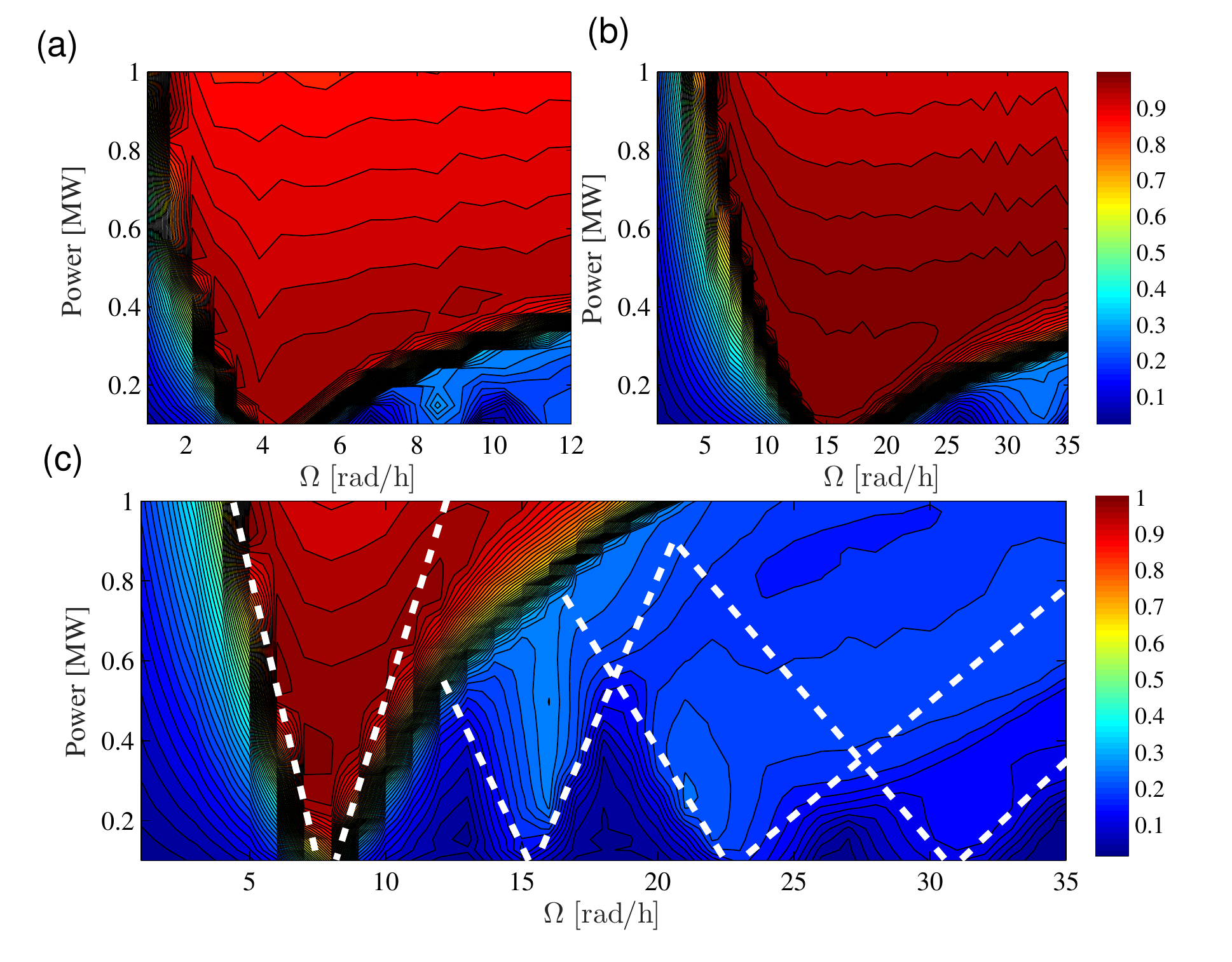}
		\caption{Arnold tongues of three different TCL ensembles of 10,000 units obtained by numerical simulations without a controller. The contour maps represent the RMS errors in tracking a sinusoidal input with frequency $\Omega$.
			(a)  The average frequency of the ensemble was $\omega_0 = 4.3$ rad/s with the thermal capacitance $C=10$ kWh/$^{\circ}$C and deadband $\delta=0.5^{\circ}$C.
			(b) The average frequency of the ensemble was $\omega_0 = 17.1$ rad/s with the thermal capacitance $C=2.5$ kWh/$^{\circ}$C and deadband $\delta=0.5^{\circ}$C.
			(c) The average frequency of the ensemble was $\omega_0 = 7.9$ rad/s with the thermal capacitance $C=1.8$ kWh/$^{\circ}$C and deadband $\delta=1.5^{\circ}$C. The white lines ($- - $)
			correspond to the theoretical Arnold tongues with whole number ratios in Fig. \ref{Fig_TheoreticalArnold_SinInput}.
			The power is normalized to 1 MW. The peak power actually corresponds to 15 MW.
		}
		\label{Fig_NumArnoldT_ThreeInOne}
		\vspace{-0ex}
	\end{figure}

	The shape and width of Arnold tongues depend on the PRC and the control input waveform. For entrainment purposes, the control input waveform can be designed to increase the width of the Arnold tongue resulting in a maximum entrainment range \cite{tanaka2015optimal} or fast entrainment \cite{zlotnik2014optimal}. While maximizing the width of the Arnold tongue is good for entrainment, for control of TCL ensembles it is to be avoided. In Fig. \ref{Fig_NumArnoldT_ThreeInOne} we have generated Arnold tongues for three different sets of TCL ensembles with different mean frequencies $\omega_0$ by measuring the Root Mean Square Error (RMSE) of the aggregate power tracking with respect to a sinusoidal reference signal with frequency $\Omega$ and amplitude $A$. The tracking error was measured at different power level $A$ and frequency $\Omega$.
	
\begin{figure}[!h]
		\centering
		\includegraphics[width= .9\linewidth]{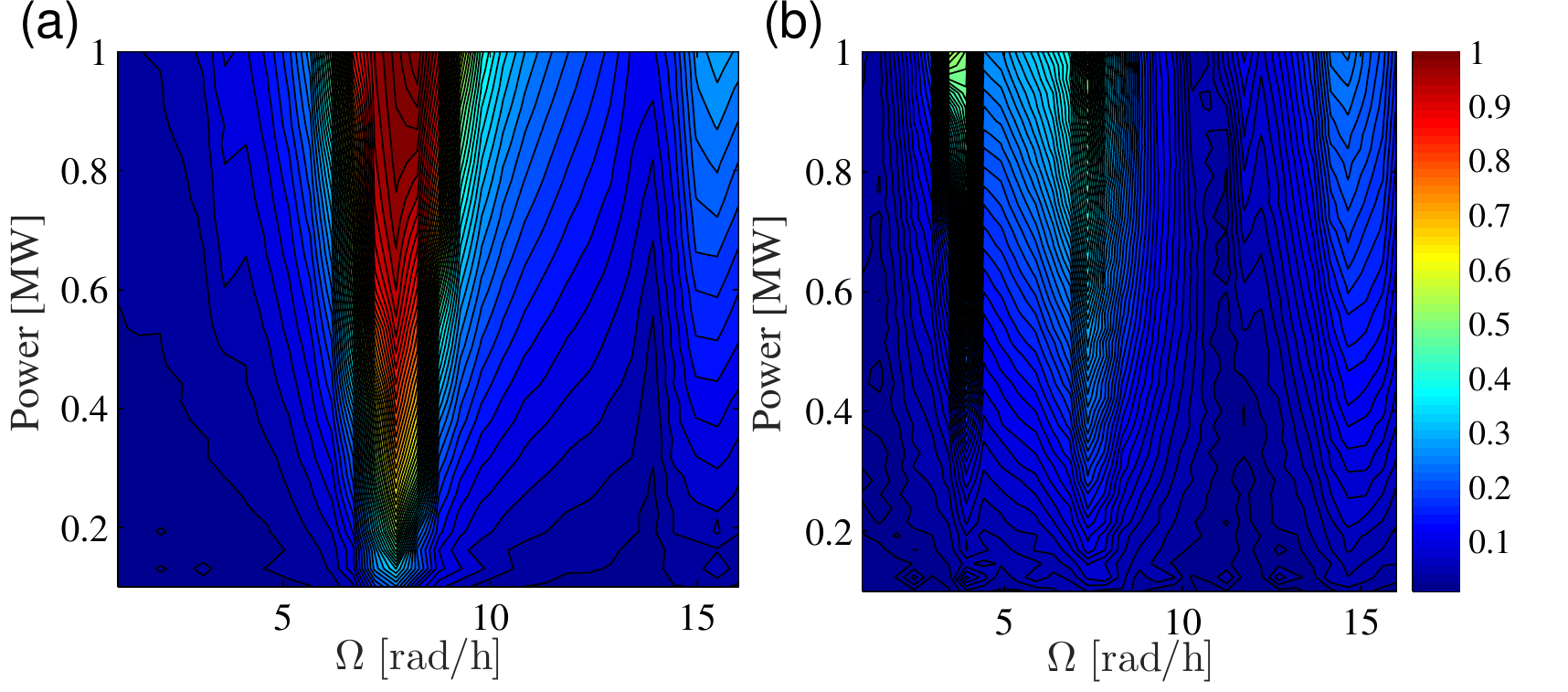}
		\caption{Arnold tongues of TCL ensembles of 10,000 units with controllers.
			(a) With PI controller. (b) With PRC-based controller. The parameter in Table 1 were used with the thermal capacitance $C=1.8$ kWh/$^{\circ}$C and the deadband $\delta=1.5^{\circ}$C.
			The power is normalized to 1. The peak power corresponds to 15 MW.
		} \label{Fig_ArnoldT_Control}
		\vspace{-2ex}
	\end{figure}

	As it can be seen from Fig. \ref{Fig_NumArnoldT_ThreeInOne}c, the Arnold tongues corresponding to the 1:1 entrainment result in the highest tracking error. The 2:1, 3:1 and 4:1 entrainment regions also have a relatively high tracking error. This is because temporary synchronization happens and there are more TCLs turning ON or OFF at the same time than it is needed for tracking a reference.
	Hence by computing the Arnold tongue one can determine the regions in the power {\em{vs.}} frequency (P,$\Omega$) space where the TCLs can provide ancillary service with minimal oscillatory response or better accuracy.
	Unlike the bound provided in \cite{barooah2015spectral}, which suggests that the tracking capacity decreases linearly with frequency, the results in this paper show that this is not entirely the case.  Fig. \ref{Fig_NumArnoldT_ThreeInOne}c for instance shows that indeed the tracking capacity decreases linearly with the input frequency and reaches its minimum at $\Omega=\omega_0$ but, it increases and decreases again forming bell shapes in the intervals $[\w_0, 2\w_0]$, $[2\w_0, 3\w_0]$ and $[3\w_0, 4\w_0]$. This implies that it is possible to extract responsive regulation from a TCL ensemble at time-scales well beyond the average of natural frequencies of the TCLs by applying an appropriate control policy that minimizes the area of the Arnold tongue.
	An example of the Arnold tongue of a TCL ensemble controlled with a proportional and integral (PI) controller is shown in Fig. \ref{Fig_ArnoldT_Control}a, and for the PRC-based controller the Arnold tongue is shown in Fig. \ref{Fig_ArnoldT_Control}b.
	
	The PI controller reduces the width of the 1:1 Arnold tongue, and the PRC-based controller further reduces it and significantly minimizes its intensity, but the Arnold tongue corresponding to the 1:2 entrainment becomes apparent. This once more confirms that it is possible to obtain significantly more demand response capacity from a given TCL population by using an appropriate controller (or control waveform).

	\subsection{Tracking of a Power Regulation Signal Based on Spectral Decomposition }
	\label{sec:Tracking}		
\begin{figure}[!h]
		\centering
		\includegraphics[width=1\linewidth]{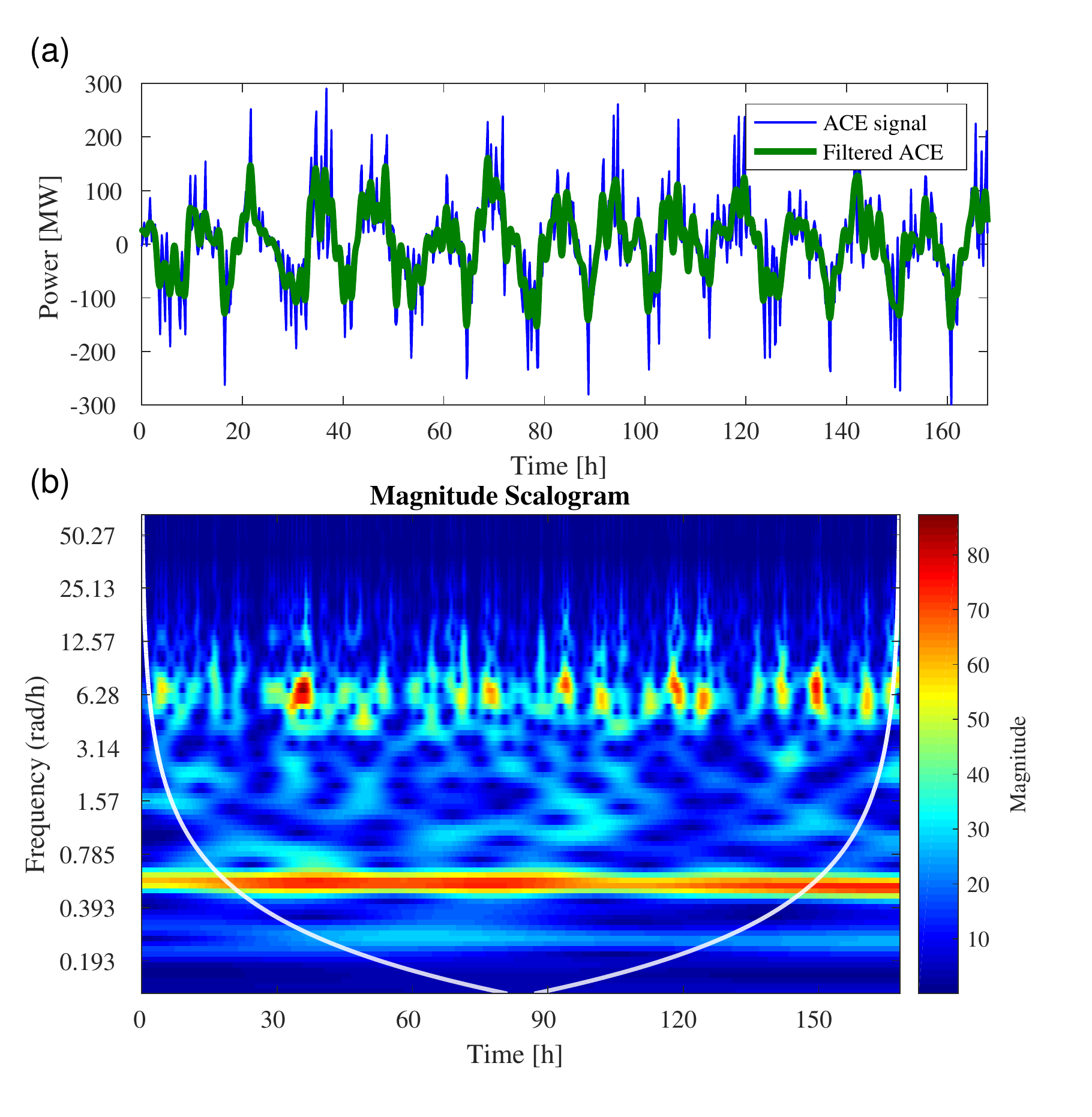}
		\caption{Area control error (ACE) signal and its power spectrum. The power spectrum was computed using wavelet transform.  (a) ACE signal with its lowpass filtered content. The cutoff frequency was 4 rad/h. (b)  ACE frequency content vs. time.
		}
		\label{Fig_PowerSpecofACE_signal}
		\vspace{-1ex}
	\end{figure}

	In this section we demonstrate the tracking of an ACE signal by using the Arnold tongues in Fig. \ref{Fig_NumArnoldT_ThreeInOne} to determine the appropriate spectral decomposition to apply to the ACE signal.
	We identify each TCL ensemble by its mean natural frequency $\omega_0$.
	The ACE signal to track is shown in Fig. \ref{Fig_PowerSpecofACE_signal}a and its wavelet transform is shown in Fig. \ref{Fig_PowerSpecofACE_signal}b, in which we can see that the dominant frequency is $\omega=0.59$ rad/h and the second dominant frequency is around $\omega = 2 \pi$ rad/h.
	
\begin{figure}[!h]
		\centering
		\includegraphics[width=1\linewidth]{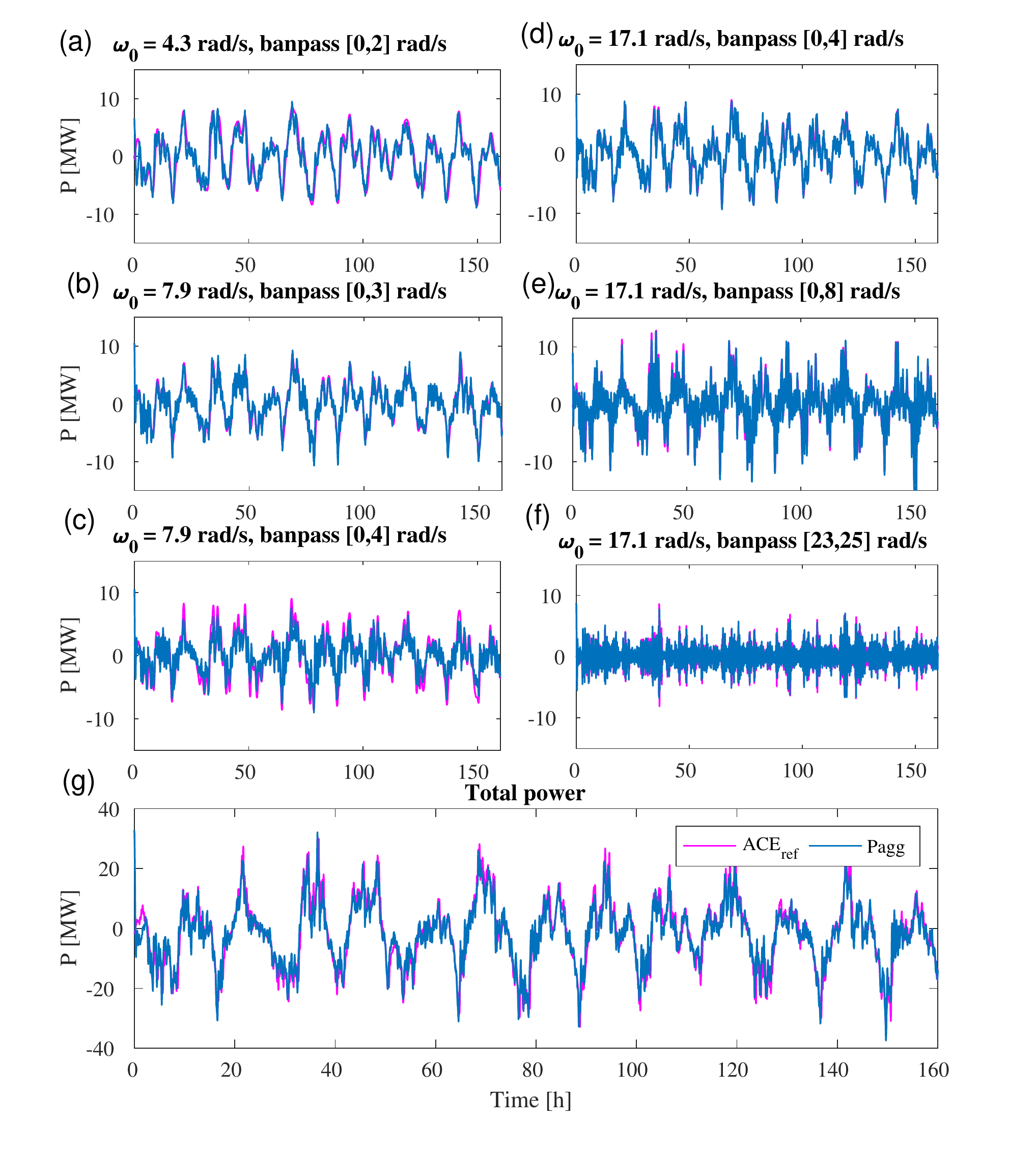}
		\caption{Tracking of the ACE by three different groups of TCL populations with 10,000 units each. The groups are identified by their mean natural frequencies $\w_0$ corresponding to specific sets of parameters $C$ and $\delta$. The rest of the parameters are the same as in Table 1.
			(a) Power tracking of an ensemble of TCL with $\w_0 = 4.3$ rad/h and the low-pass filter cutoff frequency is $\w_c=2$ rad/h.
			(b) Power tracking of an ensemble of TCL with $\w_0 = 7.9$ rad/h and the low-pass filter cutoff frequency is $\w_c=3$ rad/h.
			(c) Power tracking of an ensemble of TCL with $\w_0 = 7.9$ rad/h and the low-pass filter cutoff frequency is $\w_c=4$ rad/h.
			(d) Power tracking of an ensemble of TCL with $\w_0 = 17.1$ rad/h and the low-pass filter cutoff frequency is $\w_c=4$ rad/h.
			(e) Power tracking of an ensemble of TCL with $\w_0 = 17.1$ rad/h and the low-pass filter cutoff frequency is $\w_c=8$ rad/h.
			(f) Power tracking of an ensemble of TCL with $\w_0 = 17.1$ rad/h and the bandpass filter cutoff frequencies are $\w_l=23$ and $\w_h=25$ rad/h.
			(g) The total power $P_{\mathrm{agg}}$ is obtained as a sum of the power in (a), (b), (e) and (f).  The $ACE_{ref}$ signal is the sum of the filtered ACE signal used in (a), (b), (e) and (f) which almost recovers all the power content of the scaled down original  signal in Fig. \ref{Fig_PowerSpecofACE_signal}. Comparing $P_{\mathrm{agg}}$  and ACE shows that we are able to recover most of the power spectrum of the ACE reference power.
		}
		\label{Fig_ACEDecomposed_Tracking_AllinOne}
		\vspace{-1ex}
	\end{figure}

	The ACE signal contains different frequency components (low and high frequencies) that can be decomposed into different bands so that a certain population of TCLs with a given mean natural frequency is able to accurately track the reference power.
	In Fig. \ref{Fig_ACEDecomposed_Tracking_AllinOne}, we show different groups of TCLs tracking the ACE signal that has been filtered in specific frequency bands and scaled such that it can be tracked by the ensemble. Each TCL group is composed of $10,000$ heterogeneous units.
	For the TCLs with $\w_0 = 17.1$ rad/h, we show that this group can track low and high frequencies that are contained in the ACE signal (Figs. \ref{Fig_ACEDecomposed_Tracking_AllinOne}d, e and f). In Fig. \ref{Fig_ACEDecomposed_Tracking_AllinOne}g, we show that it is possible to nearly recover the full spectrum of the ACE signal by summing the powers that were tracked in different bands (e.g., the power Figs. \ref{Fig_ACEDecomposed_Tracking_AllinOne}a, b, e and f). Hence, by partitioning a large population of TCLs in different groups based of their capacity to track specific frequency bands, it is possible to provide ancillary services in different time scales.
	
\begin{figure}[!h]
		\centering
		\includegraphics[width=1\linewidth]{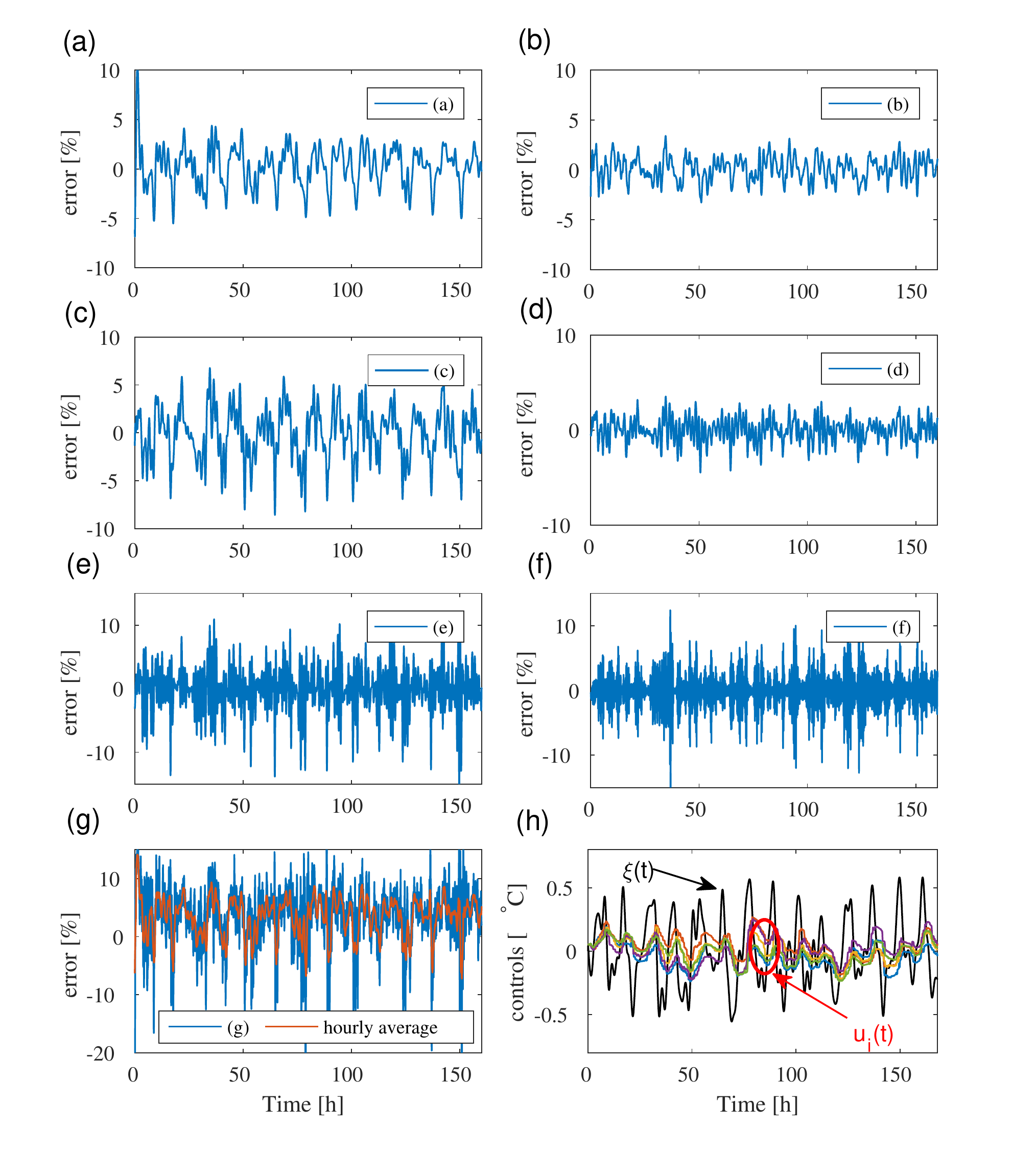}
		\caption{Relative tracking errors and sample TCL control inputs.  The legends (a) to (g) indicate the relative tracking errors in percentage of the corresponding powers in Fig. \ref{Fig_ACEDecomposed_Tracking_AllinOne} (a) to (g). The hourly average error in (g) represents the computed average of the error using a sliding window of one hour. This shows that on an hourly average the relative error is less than 10\%.  (h) Shows the global reference signal $\xi(t)$ sent by the BPA that requests of each TCL to change their power consumptions such the ACE signal can be tracked. After going through the local controllers (see Fig. \ref{fig:ControlBlockDiagram}), the TCLs control signals $u_i(t)$ are generated.
		}
		\label{Fig_ACE_TrackingError_and_Controls}
		\vspace{-1ex}
	\end{figure}

	In Fig. \ref{Fig_ACE_TrackingError_and_Controls} we show the  relative percent error of each group tracking different frequency bands.  The relative error is computed as $err(t) = (P_{ref}(t)-P_{agg}(t) )/P_{ref}(t)\times 100$\% around the baseline power $P_{base}$, which is the average power consumption of 10,000 TCLs.
	The tracking RMSE (normalized by the average aggregate power) for each case in Fig. \ref{Fig_ACEDecomposed_Tracking_AllinOne} are given in Table \ref{tab:results} where the labeling of TCL groups (a) to (g) corresponds to the labeling used in Fig. \ref{Fig_ACEDecomposed_Tracking_AllinOne} and \ref{Fig_ACE_TrackingError_and_Controls}. The RMSE is computed as
	$$\text{RMSE \%} =\sqrt{ \frac{1}{T} \frac{\int_{0}^{T} (P_{ref}(t)-P_{agg}(t) )^2 dt}{P_{base}^2 } } \times 100.$$
	
	\begin{table}[!h]
		\renewcommand{\arraystretch}{1.3}
		\caption{Results on the tracking of the ACE regulation signal }
		\label{tab:results}
		\centering
		\begin{tabular}{c|c|c|c|c|c|c|c}
			\hline
			TCL groups  &  (a) & (b)  & (c) & (d) & (e) & (f)  & (g) \\
			\hline
			\hline	
			$P_{max}$ [MW]   &  10.3 & 11.2  & 10.5 & 9.85 & 16.5 & 8.72  & 39.5  \\
			RMSE \%                &   2.31 & 1.92  & 3.09 & 1.38 & 2.92 & 3.39  & 5.89  \\	
			\hline
		\end{tabular}
	\end{table}
	The performance of the PRC-based controller is noteworthy given that communication between the BA and TCLs is open-loop. For comparison, consider for example the tracking performances in \cite{Callaway09, Vrettos2012load, Koch11}. The minimum variance control law (MVC) in \cite{Callaway09} achieves a tracking relative error of less than $5\%$ with $\text{RMSE} < 1\%$. There, the author used 60,000 TCLs (with the same power as in Table I) to track the output of a wind farm power with the 4 hours moving average subtracted from it. The resulting zero-mean power reference had a peak power of approximately 45MW but the frequency spectrum of that signal was not provided.  This is equivalent to tracking about 7.5MW with 10,000 TCLs, which is the order of magnitude that we examine here. The MVC controller used the measured aggregate power and the predicted reference power to form the control input, whereas our approach does not require the measurement of aggregate power.
	Lack of observation of aggregate power is the main justification of the increased tolerance for error in our approach, which nevertheless remains reasonably low.  If we were to use the aggregate power as feedback in our control, we could reduce the error by nearly $50\%$. For example the RMSE for the case in Fig. \ref{Fig_ACEDecomposed_Tracking_AllinOne}c is reduced to $1.8\%$ from $3.1\%$.  By incorporating $P_{agg}$ as feedback in our controller, its peak relative error is reduced to approximately $ 4\%$ from approximately $8\%$. The control is then of the form
	$$u_{k+1} = u_k+I_1 h [ I_2 \text{sgn}(\frac{1}{2}-s_k) Z(\phi_k)  (\xi_k-\xi_k^{Pagg}) - u_k ],$$
	where $\xi_k^{Pagg}$ is an appropriately scaled regulation signal that depends on the aggregate power.\
	In \cite{Vrettos2012load} four different methods were evaluated, two of which use state feedback along with the aggregate power, and the other two only use the aggregate power in the construction of the control signal. The two approaches that use state feedback achieved the RMSE of $1.18\%$ and $8.7\%$, respectively, with a maximum relative error of $30\%$. The other two controllers that do not employ any state feedback have relative errors of up $50\%$. And at last in \cite{Koch11}, the proposed feedback controller achieved an RMSE of $\leq 2.27\%$ while tracking a reference power of $4$MW with a population of 1,000 TCLs.
	This comparison shows that the results we obtained with the PRC-based approach compare well to several feedback control laws proposed in the literature in terms of reference tracking as well as minimal effect on customer comfort (see the controls in Fig. \ref{Fig_ACE_TrackingError_and_Controls}h).
	
	With this tracking example, we have demonstrated that it is possible to use TCL ensembles to provide ancillary services at different time scales by tracking low and high frequencies contained in the ACE signal. More interestingly, we arrive at the same observation as in \cite{Callaway09}, namely, that load populations with smaller thermal time constants are better candidates for providing ancillary services. We confirm this in numerical experiments and, more importantly, by computing the Arnold tongues in Fig. \ref{Fig_NumArnoldT_ThreeInOne}, which demonstrate that the loads with high natural frequency have larger power spectrum band in which they can provide ancillary services.
	
	
	\section{Conclusions and Future Directions}
	\label{sec:Conclusion}
	We have presented a new modeling paradigm for thermostatically controlled loads using neuroscience-inspired oscillator modeling and phase model reduction. The simplicity of the model opens the door to a rich variety of analysis tools and control strategies that could improve the way in which TCLs are operated to provide demand response on a power distribution system.
	Successful control policies should be able to provide demand response on time-scales of interest while satisfying the TCLs' operating constraints and customer comfort.  Specifically, the frequency at which a load is made to switch ON and OFF in a given time period should be kept within a reasonable limit to avoid excessive wear on equipment.
	
	We have developed a methodology for open-loop modulation of TCL power consumption by using phase models of the response to modulation of power, which only advances or delays the phase of the TCL power cycle when it is in a particular state, e.g., close to turning ON or turning OFF, as seen in Figure \ref{fig:PRC_p_V2}c.    We have shown that tracking of aggregate power set-points requested by a system operator could be achieved using a PRC-based control policy that prevents excessive synchronization and unwanted power fluctuations.  Moreover, we have used the concept of the Arnold tongue to characterize how well such unwanted effects are prevented.  This approach could be used to quantify the performance of control policies that do not rely on communicating feedback of the TCL states to the balancing authority.
	
	The use of simplified phase models leads to control policies that are robust to heterogeneity in system parameters.  Such models also simplify the potential formulation of optimization problems that can be solved to estimate the maximum actuation of aggregated power consumption by TCLs given state constraints and distribution subsystem size or structure.  Solution of such an optimization problem would quantify the demand response that can be obtained from a given collection of TCLs while maintaining distribution system stability and power quality.  The application of phase models to evaluate the ability of random ON/OFF switching policies to provide demand response on different time-scales is also promising.

	
	\appendix
	
	\section{TCL Model with Power and Temperature as Variables}
	\label{sec:Appendix_A}
	
	Consider the model in \eqref{eq:Continuous_mdl_v1} with the switching function defined in \eqref{eq:switch_fct} together with the aggregate power equation \eqref{eq:P_aggr}.
	Let $y(t) = 1/\eta s(t)P$ be the instantaneous power drawn by a single TCL. The time derivative of the instantaneous power is given by
	$\dot{y}(t) = 1/\eta \dot{s}(t)P$.
	The switching function in \eqref{eq:switch_fct} can also be written using an exponential function as   $s(t) = (1+\exp(-kx))^{-1}$.
	We can then write $x(t)$ in terms of the switching function as
	\begin{equation}\label{eq:x_(s)}
	x(t) = -\frac{1}{k}\ln \left(\frac{1-s}{s}\right).
	\end{equation}		
	Taking the time derivative of $x(t)$ yields
	\begin{equation*}
		-k\dfrac{dx}{dt} = \dfrac{d}{dt}(\ln(1-s)-\ln(s)).
	\end{equation*}	
	Let $u = 1-s$, by applying the chain rule we get
	\begin{equation*}  
	\begin{split}
	-k\dfrac{dx}{dt} &= \dfrac{d}{dt} \ln(u) - \dfrac{d}{dt}\ln(s),\\
	     &=\dfrac{d}{du}\ln(u) \dfrac{du}{dt} - \dfrac{d}{ds} \ln(s) \dfrac{ds}{dt},\\
	     &= \frac{1}{u}\dfrac{du}{dt}-\frac{1}{s}\dfrac{ds}{dt}.
	\end{split}
	\end{equation*}	
	After substituting for $u$ and $\dfrac{du}{dt} = -\dfrac{ds}{dt}$, we obtain
	\begin{equation}\label{eq:x_dot}
	\dot{x} = \frac{\dot{s}}{ks(1-s)}.
	\end{equation}
	We may now write \eqref{eq:x_dot} in terms of the instantaneous power $y(t)$ as
	\begin{equation}\label{eq:x_dot2}
	\dot{x} = \frac{\dot{y}}{k(1-\frac{\eta}{P}y)}.
	\end{equation}
	Substituting for \eqref{eq:x_dot2} and \eqref{eq:x_(s)} in \eqref{eq:Continuous_mdl_v1} with $s(t) = \frac{\eta}{P}y$, we obtain
	\begin{equation}
	\begin{split}
	\dot{y}(t) &= \mu k(\frac{\delta}{2}\bar{y} - \frac{1}{3} \bar{y}^3 + \t - \t_s)(1-\frac{\eta}{P}y)y,\\
	\dot{\t}(t) &= -\frac{1}{CR}(\t - \t_a + \eta R y),
	\end{split}
	\end{equation}
	where $\bar{y}(t) = -\frac{1}{k}\ln(\frac{P}{\eta y}-1)$.
	
	\section{Phase Model and Phase Response Curve}
	\label{sec:Appendix_B}
	The  phase reduction theory is a powerful tool for studying multi-dimensional rhythmic systems that are reduced to a scalar differential equation that is much easier to analyze and control.
	The autonomous oscillatory system is then described by its phase variable $\phi$ rotating on a circle $\mathbb{S}^1$.  This is represented by the phase equation $\dot{\phi}(t) = \omega$. In neuroscience the origin of the phase $\phi$ is defined as the time since the last spike of a neuron \cite{izhikevich2007dynamical}, and in our work we define it as the phase corresponding to the time the TCL turns ON.
	When an oscillator receives a pulse of strength $A$ and duration $\Delta T$, the magnitude of the induced phase shift is given by PRC$(\phi)=\phi_{new}-\phi$ \cite{izhikevich2007dynamical, Nakao15}.
	
	For completeness, we summarize the derivation of the phase model here. More details can be found in \cite{izhikevich2007dynamical, Efimov09}.
	Consider a smooth dynamical system described by 
	\begin{equation*}\label{eq:dyn_eq}
		\dot{x} = f(x,u),
	\end{equation*}
	where the state variable $x(t) \in \mathbb{R}^n$ and the input $u(t) \in \mathcal{U} \subseteq \mathbb{R}^m$. Suppose that the unforced system $\dot{x} = f(x,0)$ evolves on an attractive periodic orbit $\Gamma \subset \mathbb{R}^n$ with period $T$. The limit cycle is then described by a non-constant periodic trajectory $\gamma(t) = \gamma(t+T) \in \chi, \  \forall t \geq 0$. The linearized system along the limit cycle is given by
	\begin{equation*}
		\delta\dot{x} = A(t) \delta x(t) + B(t) u(t),
	\end{equation*}
	where  $A(t) = \frac{\partial f}{\partial x} (\g(t),0)$ and $B(t) = \frac{\partial f}{\partial u} (\g(t),0)$ are $T$-periodic.
	Given that the limit cycle $\Gamma$ is a one-dimensional closed curve \cite{izhikevich2007dynamical}, the position of any point $x_0 \in \Gamma$  can be uniquely described by a scalar phase $\phi_0 \in \mathbb{S}^1 = [0, 2\pi)$ \cite{Nakao15}.
	Let's introduce the phase function $\Theta(x)$ that maps each point $x_0$ on the limit cycle to its phase $\phi_0 = \Theta(x_0)$. The phase variable $\phi:\mathbb{R}_{\geq0} \to \mathbb{S}^1 $ is defined for each trajectory on the limit cycle as $\phi(t) = \g(t+\omega^{-1}\phi_0)$, and is periodic because of the periodicity of $\g(t)$.
	
	From the linearized model and the asymptotic phase variable, one can derive the phase-reduced model in a neighborhood of the limit cycle $\Gamma$ for sufficiently small inputs \cite{izhikevich2007dynamical, efimov2009controlling}.  By differentiating $\phi(t)$ with respect to time in the neighborhood of $\g(t)$ using the chain rule, one obtains
	\begin{align*}
		\frac{d\Theta(x)}{dt} &= \frac{\partial \Theta}{\partial x}(\g(t)) \cdot \frac{d}{dt}x(t) + \frac{\partial \Theta}{\partial x}(\g(t)) \cdot B(t)u(t),\\
		&=  \frac{\partial \Theta}{\partial x}(\g(t)) \cdot f(x)+ \frac{\partial \Theta}{\partial x}(\g(t)) \cdot B(t)u(t),\\
		&= \w + Z(\phi) \cdot B(t)u(t),
	\end{align*}
	where we have used to the fact that $\frac{\partial \Theta}{\partial x}(\g(t)) \cdot f(x) = \w$ and $Z(\phi) = \frac{\partial \Theta}{\partial x}(\g(t))$ is the phase sensitivity function also referred to as infinitesimal PRC. The input matrix function $B(t)$ depends of the differential equations describing the system, for example $B(t) = (-\mu,0)$  for the system in \eqref{eq:Continuous_mdl_v1} where the control is a perturbation of the set-point temperature $\t_s(t)$.
	
	In the following we review the main  three methods that are commonly used to compute the phase sensitivity function. These techniques are explained with great details and illustrations in \cite{izhikevich2007dynamical}.
	
	\begin{itemize}
		\item Winfree's Approach\\
		In a sufficiently small neighborhood of the limit cycle, the PRC scales linearly with respect to the strength of the pulse. Hence one can write
		$$\text{PRC}(\phi,A) \approx Z(\phi)A,$$
		where $Z(\phi) = \partial\text{PRC}(\phi,A)/ \partial A$	at $A=0$	is the linear response or sensitivity function that quantifies the small change in the instantaneous frequency caused by the weak stimulus that was applied.
		Now assume that we apply a sufficiently small stimulus $\epsilon p(t)$ and that the perturbed trajectory remains near the limit cycle attractor at all time.
		Replacing the continuous input function $\epsilon p(t)$ with the equivalent pulse train of strength $A = \epsilon p(t_n)h$, where $h$ is the time between two consecutive pulses, and $t_n$ is the timing of the $n^{th}$ pulse, one can write the Poincar{\'e} phase map as
		$$ \phi(t_{n+1})=  \{ \phi(t_n) + Z(\phi(t_n)) \epsilon p(t_n) +h \}  \quad \text{mod T},$$
		in the form
		$$ \frac{\phi(t_n+h)-\phi(t_n)}{h} = 1+Z(\phi(t_n) \epsilon p(t_n),$$
		which is a discrete version of
		\begin{equation}\label{eq:phase_mdl_w1}
		\dot{\phi}=1+\epsilon Z(\phi) \cdot p(t),
		\end{equation}	
		in the limit $h \to 0$. Note that the phase model \eqref{eq:phase_mdl_w1} is valid for any arbitrary input function $p(t)$.\
		To summarize, Winfree's approach consists of measuring the phase shift induced by a pulse train to determine the PRC.
		
		\item Kuramoto Approach\\
		Kuramoto considered the unperturbed oscillator with  $\phi(x)$ denoting the phases of points near its limit cycle attractor. Differentiating $\phi(x)$  using the chain rule yields
		$$ \frac{d\phi(x)}{dt} = \text{grad}\ \phi \cdot \frac{dx}{dt} = \text{grad}\ \phi  \cdot f(x),$$
		where grad $\phi$ is the gradient of $\phi(x)$ with respect to the state vector of the oscillator $x \in \mathbb{R}^n$.
		However, given that on the limit cycle the flow of the vector field $f(x)$ is exactly in the direction of the periodic orbit so that $ \frac{d\phi(x)}{dt} = 1$,  we obtain the important equality
		\begin{equation} \label{eq:norm_constraint}
		\text{grad}\ \phi \cdot f(x) = 1.
		\end{equation}
		By applying the chain rule to the perturbed system
	   \begin{align*} 
		\frac{d\phi(x)}{dt} &= \text{grad}\ \phi \cdot \frac{dx}{dt},\\
		&= \text{grad} \ \phi \cdot \{ f(x)+\epsilon p(t) \},\\
		&= \text{grad}\ \phi \cdot f(x) + \epsilon \text{grad} \ \phi \cdot p(t),
		\end{align*}
		and using \eqref{eq:norm_constraint}, the phase model is obtained as
		\begin{equation}\label{eq:phase_mdl_K1}
		\dot{\phi} = 1+\epsilon \text{grad}\ \phi \cdot p(t).
		\end{equation}
		Kuramoto phase model \eqref{eq:phase_mdl_K1} and Winfree's model \eqref{eq:phase_mdl_w1} are equivalent. Hence we have  $Z(\phi) = \text{grad}\ \phi$.
		
		\item Malkin's Approach\\
		Here we formally state Malkin's theorem as in \cite{izhikevich2007dynamical}.
		Suppose the unperturbed oscillator has an exponentially stable limit cycle of period $T$. Its phase evolution is described by
		\begin{equation}\label{eq:phase_mdl_M1}
		\dot{\phi} = 1+\epsilon Q(\phi)\cdot p(t),
		\end{equation}
		where $Q$  is a $T$-periodic function that is the solution to the linear ``adjoint" equation
		\begin{align}
			\dot{Q} = - \{Df(x(t))^T \}Q,\quad \text{with } Q(0) \cdot f(x(0)) = 1,
		\end{align}
		where $Df(x(t))^T$ is the transposed Jacobian of the flow $f$ at the point $x(t)$ on the limit cycle, and the normalization condition can be replaced by $Q(t) \cdot f(x(t)) = 1, \forall t.$
	\end{itemize}
	The phase models \eqref{eq:phase_mdl_M1} and \eqref{eq:phase_mdl_K1} or \eqref{eq:phase_mdl_w1} are equivalent, hence one can see that
	$$ Z(\phi) = \text{grad}\ \phi(x) = Q(\phi).$$
	The method of the adjoint was used in this paper to numerically compute the phase sensitivity function $Z(\phi)$. Examples of computer codes for this method can be found in \cite{izhikevich2007dynamical, Nakao15}.
	
	\section{Entrainment Region}
	\label{sec:Appendix_C}
	The PRC defines the synchronization properties of an oscillator and the synchronized states as fixed points of the corresponding Poincar{\'e} phase map \cite{izhikevich2007dynamical}.  	The phenomenon of entrainment by weak forcing of limit-cycle oscillators can be modeled by
	\begin{equation}\label{eq:EntrainEq1}
	\dot{\phi} = \omega+A Z(\phi)v(\Omega t),
	\end{equation}
	where $\omega$ and $\Omega$  are the natural frequencies of the oscillator, respectively, and the forcing input is $u(t) = Av(\Omega t)$, where $v$ is $2\pi$-periodic with unit energy \cite{tanaka2014optimal}.  The region of existence of a synchronized state is called Arnold tongue \cite{schaus2006response, granada2009phase}. This region of phase-locked states on $(\Omega,A)$-plane shrinks as the intensity $A$ of the stimulus approaches 0, with $\Omega$ the frequency of the stimulus.
	In general, $m:n$  entrainment occurs when $\omega/\Omega \approx n/m$ with positive relative prime integers $n$ and $m$.  This implies that the oscillator rotates exactly $m$ times while the external forcing oscillates $n$ times. Let $\Delta = \w - \frac{m}{n}\Omega$ and by formal averaging we can write \eqref{eq:EntrainEq1} as
	\begin{equation}\label{eq:EntrainEq2}
	\frac{d \psi}{dt}= \Delta + A \Gamma_{m/n}(\psi),
	\end{equation}
	where $\psi = \phi - \frac{m}{n}\Omega t$ is a slow varying phase variable, and $\Gamma_{m/n}(\psi)$ is the interaction function determined by $Z$ and $v$ as
	\begin{align*}\label{eq:InterFunction}
		\Gamma_{m/n}(\psi) &= \frac{1}{T_{ext}} \int_{0}^{T_{ext}} Z\left(\psi + \frac{m}{n}\Omega t \right) v(\Omega t) dt,\\
		& = \frac{1}{2\pi} \int_{0}^{2\pi} Z\left(\psi + m\t \right) v(n \t) d\psi,\\
		& = \frac{1}{2\pi} \langle Z\left(\psi + m\t \right), v(n\t) \rangle,
	\end{align*}
	where $T_{ext}=\frac{2\pi}{\Omega}$ is the period of the external forcing and  $\t=\frac{\Omega t}{n} \in  [0,2\pi)$. Without loss of generality, let $A=1$ and consider the $1:1$ entrainment i.e., $n=m=1$ and write $\Gamma_{m/n}(\psi)$ as simply $\Gamma(\psi)$.  It can be shown that when the condition
	\begin{equation*}
		\min \ \Gamma(\psi) < -\Delta < \max \ \Gamma(\psi),
	\end{equation*}
	is satisfied, \eqref{eq:EntrainEq2} has at least two fixed points at which $d\psi(t)/dt = 0$ holds, and one of them is stable \cite{Nakao15, tanaka2014optimal}. The interval $\Delta$ of phase locking for a fixed input strength $A$ decreases as $A \to 0$. So, for different values of $A$ and entrainment ratios $n/m$ one obtains the Arnold tongues as shown in Fig. \ref{Fig_TheoreticalArnold_SinInput}.\\

	
	\vspace{2ex}
	\section*{Acknowledgment}
	
	The authors wish to thank Scott Backhaus, Michael Chertkov, and Sean Meyn for valuable discussions.  This work was carried out under the auspices of the National Nuclear Security Administration of the U.S. Department of Energy at Los Alamos National Laboratory under Contract No. DE-AC52-06NA25396 with the support of the Advanced Grid Modeling Research Program in the U.S. Department of Energy Office of Electricity.

	
	
	
	\bibliography{TCL_BibTeX.tex}

\end{document}